\documentclass{article}

\usepackage{arxiv}
\usepackage{tabularx}
\usepackage{pdflscape}   
\usepackage{longtable}   
\usepackage{fancyhdr}    
\usepackage{enumitem}    
\usepackage[utf8]{inputenc} 
\usepackage[T1]{fontenc}    
\usepackage{hyperref}  
\usepackage{url}            
\usepackage{booktabs}       
\usepackage{amsfonts}       
\usepackage{nicefrac}       
\usepackage{microtype}      
\usepackage{lipsum}
\usepackage{graphicx}
\graphicspath{ {./images/} }
\usepackage{comment}
\usepackage{multirow}
\usepackage{amssymb}
\title{A Roadmap to Available ICS Datasets and Testbeds for Cybersecurity Research
}

\author{
 Ebtesam J. Alqahtani \\
  Department of Information and Computer Science\\
  King Fahd University of Petroleum and Minerals\\
  Dhahran 31261, PO Box 5051, Saudi Arabia\\
  \texttt{ebtesam.j.alqahtani@gmail.com} \\
   \And
 Mohammad Hammoudeh \\
  Department of Information and Computer Science\\
  King Fahd University of Petroleum and Minerals\\
  Dhahran 31261, PO Box 5051, Saudi Arabia\\
  \texttt{M.Hammoudeh@kfupm.edu.sa} \\
}

\begin{document}
\maketitle
\begin{abstract}
Industrial Control Systems (ICS) are the backbone of many critical infrastructure sectors; however, their growing level of connectivity, long lifespan and integration with the Information Technology (IT) environment introduces numerous cybersecurity challenges. The merging of Operational Technology (OT) and IT along with the deployment of Industry 4.0 technologies increases the attack surface of ICS environments, which in turn makes them more vulnerable to advanced cyber threats. Therefore, many researchers have shown interest in the field of cybersecurity of ICS. The topics of intrusion detection, anomaly detection, threat intelligence, attack simulation and resilience assessment of ICS have received much attention. Nevertheless, the development and testing of cybersecurity solutions for ICS remains to be challenging due to the lack of appropriate datasets and experimental environment.
The main objective of this paper is to provide the roadmap of existing ICS cybersecurity datasets, testbeds and digital twins. This paper presents various taxonomies along with systematic analysis of architecture, characteristics, capabilities, pros and cons of these tools. The results of the analysis demonstrate the presence of persistent problems such as lack of standardized benchmarking datasets, lack of modern attack scenarios, insufficient number of datasets based on real operational traffic and difficulty in validating artificial intelligence-driven cybersecurity solutions. In addition to summarizing current research on ICS cybersecurity datasets and testbeds, this roadmap provides the identification of research gaps and recommendations on creation of new tools.

\end{abstract}


\section{Introduction}

Operational Technology (OT) comprises programmable systems that monitor and control physical processes, including Industrial Control Systems (ICS), automation systems, transportation systems, and physical access control systems~\cite{AutomationRevolution}. As industrial environments have evolved toward highly connected, data-driven ecosystems, Artificial Intelligence (AI) has become a key enabler of predictive maintenance, process optimisation, quality inspection, energy management, and cybersecurity. Advances in Machine Learning (ML), edge computing, and cloud analytics have expanded AI applications across the industrial lifecycle while improving operational efficiency and decision-making. However, there are various difficulties related to the deployment of AI in OT, including the lack of high-quality datasets, insufficient labelled data on attacks, as well as the requirement of high reliability and safety for mission-critical systems. There are also certain limitations with respect to existing ICS datasets and testbeds which restricting the development and evaluation of trustworthy AI-based cybersecurity solutions.

In this work, we are presenting a comprehensive study that reviews and analyses ICS cybersecurity datasets, testbeds, and Digital Twins (DTs) within a unified roadmap framework, addressing the challenges associated with selecting, evaluating, and developing suitable resources for cybersecurity research. It classifies existing datasets, testbeds and DTs, analyses their characteristics and limitations, identifies common research gaps, and proposes a roadmap for developing realistic, scalable, and interoperable resources to support future AI-driven OT security research. Section 2 reviews the related work. Sections 3 and 4 discuss the categorization of the cybersecurity datasets and testbeds for ICS, respectively. Section 5 summarizes DTs for ICS cybersecurity. Based on the results presented above, Section 6 makes a comparison of datasets, testbeds, and DTs and highlights cases and ways of their application. Section 7 reveals common research gaps among the discussed resources. Section 8 proposes a roadmap towards the creation of realistic, scalable, and interoperable cybersecurity datasets, testbeds, and DTs for ICS. Section 9 formulates recommendations for the designers and users of these resources. Section 10 summarizes the paper and its key findings.

\section{Related Works}

Existing literature can be broadly grouped into two streams. The first stream of research has focused on reviewing ICS cybersecurity testbeds from different perspectives. Recent studies have examined the design, architecture, implementation, and applications of physical, virtual, and hybrid testbeds, highlighting their roles in vulnerability assessment, cyberattack experimentation, Intrusion Detection System (IDS), and cybersecurity education~\cite{Holm2015ATestbeds,Geng2019ATestbeds}. The analysis of the IIoT testbeds is performed by comparing their communication protocols, deployment strategies and design issues with practical suggestions on development of reproducible and scalable testing environments in~\cite{Zhang2024AResearch}. Another survey focused specifically on virtual ICS testbeds, reviewing their architectures, applications, and virtualisation technologies while identifying research gaps and future directions~\cite{Ekisa2024VirtualApplications}.
The second research stream has been dedicated to publicly available ICS cybersecurity datasets and discusses their data properties, attack scenarios and suitability for IDS benchmarking~\cite{Termanini2026IntrusionRubric}. There are also joint reviews of ICS testbeds and datasets to show how they complement each other in cybersecurity research and AI model evaluation~\cite{Conti2021AResearch}.

Development of DTs became one of the promising research directions for improving ICS cybersecurity. Some surveys have provided the review of the architecture, enabling technologies, security applications and adoption challenges of DTs in the context of Industry~4.0. There have been other surveys investigating the role of DTs in IDS, vulnerability assessment, cyber ranges, design of secure cyber--physical systems, and their use in combination with context modelling and machine learning for anomaly detection, predictive threat analysis and cyber defence~\cite{Mendonca2022DigitalChallenges,El-HajjTaruItapeltoTeklitGebremariam2024SystematicApplications,Mun2025ARequirements, Abraham2025TowardsApproaches}. DTs have also been investigated as security-oriented platforms that synchronise physical assets with their virtual counterparts to support monitoring, simulation, attack analysis, security validation, digital forensics, and lifecycle-wide cybersecurity~\cite{Dietz2020UnleashingSecurity}. One study further examined cybersecurity threats across the physical, communication, synchronisation, simulation, and application layers of DT architectures, highlighting the need to preserve data integrity, availability, synchronisation fidelity, and operational reliability throughout the DT lifecycle~\cite{Alcaraz2022DigitalThreats}. A recent comprehensive survey further defined DTs from a cybersecurity perspective, classified their security applications and implementation approaches, mapped them to the NIST Cybersecurity Framework, and identified key implementation challenges, open research issues, and future directions for secure DT-enabled cyber operations~\cite{Empl2025DigitalPerspectives}.

Table~\ref{tab:Current-related-peer-reviewed-work} compares representative peer-reviewed surveys on ICS cybersecurity resources. The comparison is made based on several aspects such as the scope of the review, coverage of data sets, testbeds, and DTs, whether there are taxonomies and comparative studies, identification of gaps in research, and future research directions. Such a comparison shows that the proposed review has a wider scope by integrating data sets, testbeds, DTs, taxonomies, selection criteria, and future research directions.

\begin{table*}[htbp]
\centering
\caption{Comparison of representative peer-reviewed surveys on ICS cybersecurity resources.}
\label{tab:Current-related-peer-reviewed-work}
\scriptsize
\setlength{\tabcolsep}{3pt}

\begin{tabular}{
p{0.10\linewidth}
p{0.05\linewidth}
p{0.15\linewidth}
c
c
c
c
c
c
c
c
}
\toprule
\textbf{Work} &
\textbf{Year} &
\textbf{Focus} &
\textbf{Datasets} &
\textbf{Testbeds} &
\textbf{DTs} &
\textbf{Taxonomy} &
\textbf{Comparison} &
\textbf{Research Gaps} &
\textbf{Roadmap} &
\textbf{Guidelines} \\
\midrule

~\cite{Holm2015ATestbeds}
& 2015
& ICS Testbeds
& $\times$
& $\checkmark$
& $\times$
& $\checkmark$
& $\times$
& $\checkmark$
& $\times$
& $\times$ \\

~\cite{Geng2019ATestbeds}
& 2019
& ICS Testbeds
& $\times$
& $\checkmark$
& $\times$
& $\times$
& $\times$
& $\checkmark$
& $\times$
& $\times$ \\

~\cite{Zhang2024AResearch}
& 2024
& IIoT Testbeds
& $\times$
& $\checkmark$
& $\times$
& $\checkmark$
& $\times$
& $\checkmark$
& $\times$
& $\checkmark$ \\

~\cite{Conti2021AResearch}
& 2021
& Datasets and Testbeds
& $\checkmark$
& $\checkmark$
& $\times$
& $\checkmark$
& $\blacktriangle$
& $\checkmark$
& $\times$
& $\times$ \\

~\cite{Termanini2026IntrusionRubric}
& 2026
& ICS Datasets
& $\checkmark$
& $\times$
& $\times$
& $\checkmark$
& $\times$
& $\checkmark$
& $\times$
& $\checkmark$ \\

~\cite{Mendonca2022DigitalChallenges}
& 2022
& DT
& $\times$
& $\times$
& $\checkmark$
& $\times$
& $\times$
& $\checkmark$
& $\times$
& $\times$ \\

~\cite{El-HajjTaruItapeltoTeklitGebremariam2024SystematicApplications}
& 2024
& DT
& $\times$
& $\times$
& $\checkmark$
& $\checkmark$
& $\times$
& $\checkmark$
& $\times$
& $\times$ \\

~\cite{Mun2025ARequirements}
& 2025
& DT
& $\times$
& $\times$
& $\checkmark$
& $\checkmark$
& $\times$
& $\checkmark$
& $\times$
& $\times$ \\

\textbf{This Work}
& \textbf{2026}
& \textbf{ICS Cybersecurity Resources}
& $\checkmark$
& $\checkmark$
& $\checkmark$
& $\checkmark$
& $\checkmark$
& $\checkmark$
& $\checkmark$
& $\checkmark$ \\
\bottomrule
\end{tabular}

\begin{flushleft}
\footnotesize
\textit{Note:} $\checkmark$ = Covered; $\times$ = Not covered; $\blacktriangle$ = Partially covered. 
\end{flushleft}
\end{table*}

\section{ICS Cybersecurity Datasets}

 
ICS datasets support the development and evaluation of monitoring, control, and security solutions. This survey focuses specifically on publicly available \textit{cybersecurity-related} datasets developed for ICS. These datasets are widely used to develop and evaluate intrusion detection, anomaly detection, attack classification, and other AI-driven cybersecurity techniques. They span multiple industrial sectors, including water, power systems, manufacturing, oil and gas, transportation, and IIoT. The following sections review representative datasets according to their application domain.

Although not developed for ICS, traditional IDS datasets remain widely used as baseline benchmarks for evaluating ML models. DARPA 1998 simulated a military network with realistic background traffic and multiple attack scenarios, while KDD Cup 1999 was derived from the DARPA environment and provided labelled network intrusion data covering Denial of Service (DoS), probing, and unauthorised access attacks~\cite{LippmannResultsEvaluation, 1998Laboratory,KDDData}. NSL-KDD addressed the limitations of KDD Cup 1999 by removing redundant records and balancing attack distributions~\cite{NSL-KDDUNB}. More recent datasets include UNSW-NB15, which contains modern network traffic generated from real and synthetic attacks, and CICIDS2017, which provides labelled benign and malicious traffic captured from a realistic enterprise network environment~\cite{Moustafa2015UNSW-NB15:Set, TheResearch,IDSUNBb, Panwar2019ImplementationWEKA}.

The water industry is represented by several widely used datasets developed from realistic water storage, treatment, and distribution testbeds. WUSTL-IIoT-2018 was collected from a water storage tank testbed and includes network traffic and sensor data captured during normal operation and cyberattacks. The dataset is commonly used for anomaly detection and IIoT security research~\cite{WUSTL-IIOT-2018Research, WUSTL-IIOT-2018DataPort}. 
SWaT was collected from a real water treatment testbed and contains multivariate process data under both normal and attack conditions. It became one of the most widely used datasets for evaluating cyber-physical attack detection methods. The Water Distribution Testbed (WADI) extended the SWaT environment to a larger water distribution system and focused on analysing cascading effects of cyber-physical attacks. Moreover, new industrial communication protocols have been introduced compared to SWaT~\cite{SWaTDetection, ITrustITrust}.
BATADAL dataset was created for the Battle of the Attack Detection Algorithms contest and consists of realistic water distribution data annotated with cyberattacks. It has been actively used to compare various anomaly detection and cyber-attack detection techniques in water infrastructure~\cite{BATADAL}.

There are several datasets created to be used for cyber-physical security research within the power systems industry. The Power System Attack Dataset developed by Mississippi State University (MSU) and Oak Ridge National Laboratory (ORNL) includes data simulated cyber-attacks on components of the power grid along with the data on network traffic, system logs, and operational measurements. The dataset was intended to be used to research security of smart grids and cyber-physical attacks~\cite{PowerState}.
Electrical Power and Intelligent Control (EPIC) included data from generation, transmission, smart home, and microgrid scenarios under normal operating conditions. The dataset was used to study anomaly detection and interactions between interconnected industrial systems~\cite{Adepu2019EPIC:Security}.
Protocol-specific datasets based on IEC 61850 and DNP3 capture cyberattacks targeting substation automation and SCADA communications. These datasets are widely used to evaluate IDS and protocol-aware security mechanisms for power system applications~\cite{NetworkAttacks, DNP3}.
The PNNL High-Fidelity Cyber-Physical Dataset was developed to address the shortage of realistic OT datasets for intrusion detection research by leveraging the HAI testbed. It was specifically designed to provide a benchmark for evaluating OT-specific IDS under realistic operational conditions
~\cite{Ashok2021ASystems}.

The oil and gas industry is represented by datasets developed to capture cyber-physical attacks against PLC-controlled pipeline operations. The Gas Pipeline Dataset contained telemetry data and attack scenarios such as command injection and data injection targeting PLC-controlled pipeline operations. It was used to investigate cyber-physical attacks in oil and gas infrastructure~\cite{Beaver2013AnCommunications, IMPACTDatasets}.

Railway cybersecurity research is supported by datasets representing SCADA- and PLC-based railway environments. The Electra Dataset simulated a railway electric traction substation controlled by SCADA and PLCs communicating through protocols such as Modbus and S7Comm. It was used to evaluate ML and deep learning approaches for industrial anomaly detection~\cite{ElectraDataset, Gomez2019OnSystems}.
The NCL Railway Cyber Range dataset was generated from a simulated railway ICS environment designed to emulate modern cyberattacks against operational technology. It includes historical and emerging attack scenarios targeting railway control systems~\cite{Yusof2025SignalsRange}.
The Security Threats in Smart Railway Systems (STSRS) is a high-fidelity time-series dataset generated from a simulated smart railway Cyber-Physical Systems (CPS). It supports anomaly detection, cyber-defence modelling, and DTs evaluation for railway ICS environments~\cite{Abukeshek2026STSRS:Systems}.


Manufacturing industry is represented by ICS-Flow, which was generated from the ICSSIM bottle-filling factory simulation and included process data and attack scenarios involving PLCs, HMIs, and Modbus communication. The dataset supported research on manufacturing system security and attack detection~\cite{Dehlaghi-GhadimAnomalySystems}.

Building Automation Systems (BAS) datasets primarily focus on BACnet communications under normal and attack conditions. These datasets support the evaluation of intrusion detection and protocol-aware security mechanisms for smart buildings and related critical infrastructure~\cite{BACnetInternationalIntroductionBACnet, BACnetDataset}.

The IIoT industry is represented by several datasets. Ton\_IoT combined network traffic, operating system logs, and IoT telemetry collected from edge, fog, cloud, and IoT environments. The dataset represented modern IIoT deployments and included a wide range of attack scenarios~\cite{Booij2022ToN_IoT:Sets, ToN_IoTDataPort}.
Edge-IIoTset contains realistic IIoT network traffic generated from multiple IoT devices, communication protocols, and cyberattacks. The dataset supports the development and evaluation of AI-based IDSs~\cite{Ferrag2022Edge-IIoTset:Learning}. 
X-IIoTID provides network traffic collected from a realistic IIoT testbed under normal and attack conditions. It was designed to benchmark intrusion detection methods in modern industrial IoT environments~\cite{Al-Hawawreh2022X-IIoTID:Things}.
MQTTset (MQTT-IoT-IDS2020) focuses on MQTT-based IoT communications and includes benign and malicious traffic for evaluating intrusion detection techniques targeting MQTT-enabled IIoT environments~\cite{MQTT-IoT-IDS2020:DataPort}.
CICIoT2023 contains large-scale network traffic generated from heterogeneous IoT devices under normal operation and multiple cyberattacks. The dataset is widely used for benchmarking ML and deep learning-based IDSs~\cite{Neto2023CICIoT2023:Environment}.

ICS Vulnerability Datasets, such as ICS-LTU2022 which is a publicly available vulnerability dataset that compiles known vulnerabilities affecting ICS from multiple public sources. It provides structured information, including vulnerability identifiers, affected vendors and products, severity scores, and vulnerability descriptions, making it suitable for vulnerability analysis, risk assessment, and vulnerability prediction rather than intrusion or anomaly detection~\cite{Alanazi2025ICS-LTU2022:Vulnerabilities}.

Cross-Industry ICS datasets include ICS-ADD, which was generated from an open-source ICS testbed. It provides security events generated by OSSIM and Suricata, making it suitable for intrusion detection, anomaly detection, and security monitoring research~\cite{Gaggero2024IndustrialEnvironments}.
Cyber4OT contains network traffic collected from a realistic ICS laboratory testbed. The testbed incorporates PLCs, SCADA, HMI, industrial and office networks, and supports multiple communication protocols. The dataset is intended for intrusion detection, vulnerability assessment, and AI-based cybersecurity research in ICS~\cite{Cabaj2025Cyber4OTSystems}.


Table~\ref{tab:dataset_comparison} summarises all ICS cybersecurity datasets reviewed in this study. The datasets are categorised according to their application domain and compared based on the type of data collected, dataset source, availability of labelled attack data, and their primary ML application. 

\begin{table*}
\caption{Comparison of Representative ICS Cybersecurity Datasets}
\label{tab:dataset_comparison}
\centering
\scriptsize
\begin{tabular}{
p{0.10\textwidth}
p{0.10\textwidth}
p{0.12\textwidth}
p{0.11\textwidth}
p{0.04\textwidth}
p{0.15\textwidth}
p{0.10\textwidth}
p{0.07\textwidth}
}
\toprule
\textbf{Category} & \textbf{Dataset} & \textbf{Data Type} & \textbf{Collection Source} & \textbf{Labels} & \textbf{Attack} & \textbf{Protocols} & \textbf{Primary Use} \\
\midrule

\multirow{5}{*}{Traditional IDS}
& DARPA 1998 & Network & Simulated & \checkmark & DoS, Probe, R2L, U2R & TCP/IP & IDS \\
& KDD Cup 99 & Network & Simulated & \checkmark & DoS, Probe, R2L, U2R & TCP/IP & IDS \\
& NSL-KDD & Network & Simulated & \checkmark & DoS, Probe, R2L, U2R & TCP/IP & IDS \\
& UNSW-NB15 & Network & Hybrid & \checkmark & DoS, Exploits, Fuzzing, Worms & TCP/IP & IDS \\
& CICIDS2017 & Network & Realistic & \checkmark & DoS, DDoS, Botnet, Web Attacks & TCP/IP & IDS \\
\midrule

\multirow{4}{*}{Water}
& SWaT & Process + Network & Physical & \checkmark & FDI, Replay, DoS & Ethernet/IP & AD \\
& WADI & Process + Network & Physical & \checkmark & FDI, Replay, DoS & Modbus/TCP & AD \\
& BATADAL & Process & Simulated & \checkmark & FDI & SCADA & AD \\
& WUSTL-IIoT-2018 & Sensor + Network & Hybrid & \checkmark & DoS, MITM, FDI & MQTT, Modbus/TCP & IDS/AD \\
\midrule

\multirow{6}{*}{Power Systems}
& Power System Attack & Process + Network & Physical & \checkmark & FDI, Replay, DoS & DNP3 & IDS/AD \\
& EPIC & Process & Physical & \checkmark & FDI, Replay & IEC 61850, DNP3 & AD \\
& IEC 61850 Datasets & Network & Physical & \checkmark & DoS, MITM, GOOSE Injection & IEC 61850 & IDS \\
& DNP3 Datasets & Network & Physical & \checkmark & Replay, MITM, DoS & DNP3 & IDS \\
& PNNL High-Fidelity & Process + Network & Hardware-in-the-Loop & \checkmark & Reconnaissance, MITM, Spoofing, Fuzzing, Data Injection, DoS & DNP3, Modbus/TCP, BSAP & IDS/AD \\
\midrule

Oil \& Gas
& Gas Pipeline & Process & Physical & \checkmark & Command Injection, FDI & Modbus/TCP & AD \\
\midrule

\multirow{3}{*}{Railway}

& Electra & Process + Network & Simulated & \checkmark & FDI, Replay, DoS & Modbus/TCP, S7Comm & AD \\
& NCL Railway Cyber Range &Process + Network & Simulated & \checkmark & DoS, Replay, FDI, MITM & Modbus/TCP, IEC 60870-5-104 & IDS/AD\\
& STSRS & Time-series + Network & Simulated & \checkmark & DoS, Replay, Jamming & TCP/IP & IDS/AD \\
\midrule

Manufacturing
& ICS-Flow & Process + Network & Simulated & \checkmark & FDI, Replay, DoS & Modbus/TCP & IDS \\
\midrule

Building Automation
& BACnet & Network & Physical & \checkmark & DoS, Replay, MITM & BACnet/IP & IDS \\
\midrule

\multirow{5}{*}{IIoT}
& ToN-IoT & Mixed & Hybrid & \checkmark & DoS, Backdoor, Ransomware, Injection & MQTT, HTTP, Modbus/TCP & IDS \\
& Edge-IIoTset & Network & Physical & \checkmark & DoS, MITM, Malware, Injection & MQTT, CoAP, Modbus/TCP & IDS \\
& X-IIoTID & Network & Physical & \checkmark & DoS, FDI, MITM & MQTT, Modbus/TCP & IDS \\
& MQTTset & MQTT Traffic & Physical & \checkmark & Flooding, DoS, Brute Force & MQTT & IDS \\
& CICIoT2023 & Network & Physical & \checkmark & DoS, DDoS, Botnet, Spoofing & MQTT, CoAP, HTTP & IDS \\
\midrule

\multirow{2}{*}{Cross-Industry ICS}
& ICS-ADD & Network + Security Events & Physical & \checkmark & DoS, MITM, Malware & Modbus/TCP & IDS/AD \\
& Cyber4OT & Network + Process & Physical & \checkmark & DoS, Replay, MITM & Modbus/TCP, TCP/IP & IDS/AD \\
\midrule

Vulnerability Datasets
& ICS-LTU2022 & Vulnerability Records & Public Repositories & \checkmark & Vulnerabilities (CVE/CWE) & N/A & Vulnerability Analysis\\
\bottomrule
\end{tabular}
\begin{flushleft}
\footnotesize{\textit{Note:} IDS = Intrusion Detection System; AD = Anomaly Detection; FDI = False Data Injection; MITM = Man-in-the-Middle; DoS = Denial of Service; R2L = Remote-to-Local; U2R = User-to-Root.}
\end{flushleft}
\end{table*}

\section{ICS Cybersecurity Testbeds}

While public benchmarks are useful for comparison purposes, their static nature makes it impossible to conduct any controlled experiments with them, as well as data generation. As such, ICS testbeds have been adopted by many organizations for cybersecurity testing.



Physical ICS testbeds integrate real industrial hardware and communication networks to provide high-fidelity environments for cybersecurity experimentation. It enables realistic attack execution, defence validation, and dataset generation while closely reflecting operational industrial environments.
The Sam Houston State University (SHSU) SCADA Testbed is a low-cost laboratory developed to support industrial cybersecurity research and digital forensics. It provides a realistic environment for penetration testing, vulnerability assessment, incident response, and forensic investigations~\cite{Krishnan2019SCADAForensics}.
CrossTest is a cross-domain physical ICS cybersecurity testbed that combines energy and manufacturing systems to support the evaluation of domain-independent threat detection techniques. It includes multiple cyberattack scenarios and publicly available network traffic datasets~\cite{Karch2022CrossTest:Evaluations}.
SPHERE CPS Enclave is a modular, remotely accessible ICS testbed designed to support cybersecurity experimentation on PLCs, industrial networks, and DTs. It provides a configurable environment for investigating cyber-physical attacks, anomaly detection, intrusion resilience, and Hardware-In-the-Loop (HIL) validation, enabling controlled and reproducible cybersecurity experiments~\cite{Garcia2025SPHEREExperimentation}.
The University of New Orleans (UNO) SCADA Testbed is a laboratory-scale ICS testbed that emulates gas pipeline, power distribution, and wastewater treatment processes using industrial hardware. It supports multiple industrial communication protocols and enables realistic cyberattack experimentation, security evaluation, and defence validation across several industrial sectors~\cite{Ahmed2016APedagogy}.

Virtual testbeds replicate industrial control systems through simulations in software form, offering an efficient platform for experimenting in the field of cybersecurity without necessitating expensive industrial hardware. 
The VICSORT is an open-source virtualised ICS testbed built upon the Graphical Realism Framework for Industrial Control Systems (GRFICS). It provides a realistic yet lightweight environment for cybersecurity education, attack simulation, defence validation, and security evaluation while simplifying deployment and reducing hardware requirements~\cite{Ekisa2022VICSORT-ATestbed}.
The Open Virtual Testbed Platform is an open-source virtual ICS framework designed to support reproducible cybersecurity research, intrusion detection evaluation, and dataset generation while allowing interoperability with real industrial devices~\cite{Reaves2012AnResearch}.
The Virtual Tennessee-Eastman Testbed (VTET) is a virtual ICS testbed based on the Tennessee-Eastman chemical process. It supports both fully virtual and semi-virtual deployments, integrates physical or virtual PLCs, and implements multiple industrial communication protocols for cybersecurity experimentation~\cite{Xie2018VTET:Research}.
Cyber Agent Evaluation Testbed is a configurable OT network emulation platform that supports cybersecurity experimentation across power grid and chemical plant environments. Built using OpenStack virtualisation and SCADA visualisation, it enables realistic evaluation of cyber agents and attack scenarios without requiring operational industrial infrastructure~\cite{Raj2025AutonomousNetworks}.


The HIL testbeds combine real industrial hardware with real-time simulation to achieve high-fidelity cyber-physical experimentation while maintaining flexibility and safety.
The HIL Augmented ICS is a realistic HIL testbed developed to generate cybersecurity datasets for AI-based anomaly detection research. Successive releases expanded monitored variables and attack scenarios, while the HAICon competition established benchmark results for AI-based anomaly detection methods, making HAI one of the most widely adopted ICS cybersecurity benchmarks~\cite{ShinHAIDataset,Shin2021TwoTestbed}.
HIL-RESIST is a modular HIL testbed for power system cybersecurity that combines real-time simulation, hardware controllers, communication emulators, and cyberattack injection to support cyber-physical experimentation, resilience evaluation, and cybersecurity validation in smart grids~\cite{Chhokra2025HIL-RESIST:Systems}.
The Testbed for Resilient Operational Systems (TROY) integrates a Typhoon-HIL real-time simulator with a virtual IT/OT environment to emulate critical infrastructure systems for cybersecurity research, resilience assessment, and workforce training~\cite{Sutterfield2024Real-TimeSystems}.
The WAMS Cyber-Physical Testbed is a HIL platform developed for wide-area monitoring system cybersecurity research. It combines commercial hardware, real-time digital simulation, and industrial communication protocols to model cyberattacks while generating labelled datasets from heterogeneous sensor data~\cite{Adhikari2017WAMSMining}.


Cloud- and container-based testbeds improve scalability, portability, and reproducibility by virtualising industrial environments using container technologies or cloud infrastructures.
The Container-based SCADA Testbed is a Docker-based virtual ICS environment that enables rapid deployment of realistic SCADA systems while generating labelled network traffic datasets for ML-based intrusion detection research~\cite{Khan2020LightweightSystems}.
The Towards a Container-Based ICS Testbed (TRIST) is a lightweight container-based virtual ICS testbed that supports reproducible and cost-effective cybersecurity experimentation by enabling rapid deployment of realistic cyber-physical environments without requiring dedicated hardware~\cite{Lo2025TRIST:Detection}.
The Cloud-Ready Virtual OPC UA Testbed is a fully virtual, Docker-based ICS platform built using OpenPLC and OPC UA to support flexible cybersecurity experimentation, attack simulation, and defence validation in modern OT environments~\cite{Giehl2024EmuFlex:UA}.
The OT Industrial Control Network Testbed emulates realistic industrial control networks by integrating engineering workstations, HMIs, PLCs, MQTT brokers, and industrial switches within a virtualised environment. It supports cyberattack experimentation while generating high-fidelity datasets using centralised monitoring through Elastic SIEM, Zeek, NetFlow, and Auditd~\cite{Rahmani2025AOT-Networks}.


Domain-specific testbeds are designed to emulate the operational characteristics of particular industrial sectors or communication technologies, enabling realistic evaluation of sector-specific cyber threats and defence mechanisms.
LICSTER is a low-cost open-source ICS testbed that enables affordable cybersecurity experimentation using commercially available hardware for education and research~\cite{Sauer2019LICSTERResearch}.
vWaterLabs is an entirely virtualized testbed based on the water treatment system process and ANSI/ISA-99 architecture. The testbed offers reproducible laboratory experiments with regard to PLC programming, industrial protocols, vulnerability assessment, and cyberattack testing~\cite{Steiner2021VWaterLabs:Education}.
The Naval Defence ICS Testbed emulates representative warship control systems, enabling realistic attack simulation, cyber defence evaluation, and representative dataset generation for AI-based cybersecurity research~\cite{Sicard2022AnResearch}.
Security Water Processing (SWaP) is a physical water treatment ICS testbed supporting industrial protocol evaluation, cyberattack experimentation, defence validation, and public dataset generation~\cite{Calder2023SWaP:Research}.
The ToN\_IoT Testbed emulates realistic IIoT environments using heterogeneous devices to generate comprehensive cybersecurity datasets containing network traffic, telemetry, system logs, IDS alerts, and threat data for intrusion detection and digital forensics research~\cite{Mekala2022IndustrialForensics}.
The Power SCADA Testbed models electrical generation and distribution substations using IEC 60870-5-104 and IEC 61850 communication protocols for power system cybersecurity research~\cite{Jarmakiewicz2015DevelopmentInfrastructure}.
The CENTER Water Management Testbed supports cybersecurity research, training, and attack–defence evaluation for waste and potable water systems by emulating realistic treatment and distribution processes using industrial protocols~\cite{Ozcelik2021CENTERManagement}.
The GOOSE Protocol Testbed focuses on evaluating the implementation and security of the IEC 61850 GOOSE protocol, enabling protocol validation, traffic generation, and mitigation of communication-layer vulnerabilities~\cite{Boeding2023AImplementations}.


Table~\ref{tab:testbed_comparison} provides a summary of the ICS testbeds examined in this study. The testbeds are classified based on their architecture and are evaluated in terms of their implementation method, communication protocol support, cybersecurity applications, and data generation features.
\begin{table*}[t]
\caption{Comparison of Representative ICS Cybersecurity Testbeds}
\label{tab:testbed_comparison}
\centering
\scriptsize
\begin{tabular}{
p{0.10\textwidth}
p{0.10\textwidth}
p{0.10\textwidth}
p{0.06\textwidth}
p{0.12\textwidth}
p{0.15\textwidth}
p{0.10\textwidth}
p{0.06\textwidth}
}
\toprule
\textbf{Category} &
\textbf{Testbed} &
\textbf{Domain} &
\textbf{Fidelity} &
\textbf{Protocols} &
\textbf{Primary Purpose} &
\textbf{Dataset Generation} &
\textbf{Availability} \\
\midrule

\multirow{4}{*}{Physical}
& SHSU SCADA & Generic SCADA & Medium & Modbus/TCP, Ethernet/IP & Digital forensics, penetration testing & $\times$ & $\times$ \\
& CrossTest & Energy + Manufacturing & High & Modbus/TCP, OPC UA & Cross-domain threat detection & \checkmark & \checkmark \\
& SPHERE CPS Enclave & Generic CPS & High & Modbus/TCP, OPC UA, DNP3 & CPS security experimentation & $\times$ & $\times$ \\
& UNO SCADA & Gas, Power, Water & High & Modbus/TCP, EtherNet/IP, PROFINET & Cross-domain attack experimentation & $\times$ & $\times$ \\
\midrule

\multirow{4}{*}{Virtual}
& VICSORT & Generic ICS & Medium & Modbus/TCP, DNP3 & Security research, education & $\times$ & \checkmark \\
& Open Virtual Testbed Platform & Generic ICS & Medium & Modbus/TCP, Ethernet/IP & IDS research & \checkmark & \checkmark \\
& VTET & Chemical Process & High & OPC, Modbus/TCP, Siemens S7 & Attack experimentation & $\times$ & $\times$ \\
& Cyber Agent Evaluation Testbed & Power + Chemical & High & Modbus/TCP, DNP3, OPC UA & Cyber-agent evaluation & $\times$ & $\times$ \\
\midrule

\multirow{4}{*}{HIL}
& HAI & Generic ICS & High & Modbus/TCP & AI anomaly detection & \checkmark & \checkmark \\
& HIL-RESIST & Smart Grid & High & IEC 61850, DNP3 & Cyber resilience evaluation & $\times$ & $\times$ \\
& TROY & Critical Infrastructure & High & Modbus/TCP, DNP3, OPC UA & Resilience testing, education & $\times$ & $\times$ \\
& WAMS Cyber-Physical & Power Systems & High & IEC 61850, DNP3 & Dataset generation, cyber-power analysis & \checkmark & $\times$ \\
\midrule

\multirow{4}{*}{Cloud/Container}
& Container-based SCADA & Generic SCADA & Medium & Modbus/TCP & Dataset generation, IDS & \checkmark & \checkmark \\
& TRIST & Generic ICS & Medium & Modbus/TCP, OPC UA & Threat simulation & \checkmark & \checkmark \\
& Cloud-Ready Virtual OPC UA & Generic ICS & Medium & OPC UA & Attack simulation, defence validation & $\times$ & \checkmark \\
& OT Industrial Network & Generic ICS & High & MQTT, Modbus/TCP, OPC UA & AI dataset generation & \checkmark & $\times$ \\
\midrule

\multirow{9}{*}{Domain-Specific}
& LICSTER & Generic ICS & Medium & Modbus/TCP & Education, attack experimentation & $\times$ & \checkmark \\
& vWaterLabs & Water & Medium & Modbus/TCP & Cybersecurity education & $\times$ & \checkmark \\
& Naval Defence ICS & Maritime & High & Modbus/TCP, OPC UA & Cyber defence evaluation & \checkmark & $\times$ \\
& SWaP & Water Treatment & High & Modbus/TCP, EtherNet/IP & Protocol security, IDS & \checkmark & $\times$ \\
& ToN\_IoT Testbed & IIoT & High & MQTT, Modbus/TCP, HTTP & Dataset generation & \checkmark & $\times$ \\
& Power SCADA Testbed & Power Grid & High & IEC 61850, IEC 60870-5-104 & Protocol security evaluation & $\times$ & $\times$ \\
& CENTER Water Testbed & Water Management & High & DNP3, Modbus/TCP, Siemens S7, EtherCAT & Training, attack--defence analysis & $\times$ & $\times$ \\
& GOOSE Protocol Testbed & IEC 61850 & High & IEC 61850 GOOSE & Protocol security evaluation & $\times$ & $\times$ \\
\bottomrule
\end{tabular}

\begin{flushleft}
\footnotesize
\textit{Note:} Fidelity reflects how closely the testbed replicates operational industrial environments. High = real industrial hardware or HIL systems; Medium = realistic virtual or emulated environments; Low = simplified simulation-based environments. Dataset Generation indicates whether the testbed supports the generation of cybersecurity datasets. Availability indicates whether the testbed is publicly available or open source.
\end{flushleft}
\end{table*}

\section{Digital Twins for ICS Cybersecurity}
DTs provide dynamic virtual representations of physical industrial systems, enabling real-time monitoring, simulation, and cybersecurity evaluation. To facilitate comparison, the reviewed DT platforms are categorised according to their primary cybersecurity purpose. 


Operational Monitoring and Simulation is enabled by a DT based on IoT for industrial motors which continuously monitors operational parameters like temperature, vibrations, and sound to identify any faults and activate protective measures automatically. Real-time visualisation, remote monitoring and predictive maintenance can be carried out by integrating the system in the cloud~\cite{Sankar2025DigitalControl}. Another DT models an industrial cocoa production process by integrating PLCs, HMIs, and IoT technologies to enable real-time monitoring, process control, and interactive simulation of industrial operations for engineering education~\cite{Ortiz2025Digital4.0}. A Cyber Twin extends conventional DTs through semantic machine models, bidirectional communication, simulation, and control to support cost-efficient Industry~4.0 application development and virtual application testing without interrupting production~\cite{Bamunuarachchi2021DigitalTransformation}.

In terms of security testing and validation, a DT for industrial networks provides an automated cyber range for cybersecurity analysis, attack experimentation, and validation of security countermeasures without affecting the physical CPS~\cite{Grasselli2022AnSystems}. A security-oriented DT of an industrial filling plant supports process-aware cyberattack simulation, synthetic dataset generation, and evaluation of ML-based IDSs using attacks such as command injection, DoS, and measurement manipulation~\cite{Varghese2022DigitalSystems}. 
Similarly, a Testing and Simulation (TaS) framework is presented to maintain real-time synchronisation with physical IoT systems to support software validation, fault injection, cybersecurity experimentation, and what-if analysis for evaluating system resilience under diverse operational and attack scenarios~\cite{Nguyen2022DigitalTool}. An automated Fuzz Testing DT framework further enables systematic vulnerability discovery, security verification, and assessment of unintended system behaviour~\cite{Dauphinais2023AutomatedSystems}. For smart grid security, a DT framework has been developed to support the complete lifecycle of smart grids, enabling continuous penetration testing and cybersecurity evaluation~\cite{Atalay2020AStandardization}. A hybrid DT for smart grids combines digital simulation with physical single-board computers to emulate CPS components and communication networks with high fidelity, proposed in~\cite{Ayyalusamy2022HybridAnalysis} to provide a scalable and cost-effective platform for cybersecurity experimentation and vulnerability assessment. A DT-based framework for IoT-enabled critical national infrastructure employs virtual replicas of network and control logic to support cybersecurity analysis while investigating blockchain and trustless security mechanisms for protecting interconnected industrial systems and critical infrastructure~\cite{Hammoudeh2020BlockchainInfrastructure}.


AI-enabled DTs integrate digital twin technology with intelligent analytics to strengthen ICS cybersecurity. One framework uses simulated operational data generated by a DT to train ML models offline and deploy them on edge devices for real-time detection of faults and cyber-attacks~\cite{Allison2022DigitalApplications}. A DT-supported anomaly detection framework combines real-time sensor monitoring with knowledge graphs and ontology-based data modelling to identify abnormal behaviour through ML~\cite{Chukkapalli2021Cyber-PhysicalUsecase}. Building on predictive analytics, the DT4I4-Secure framework combines physics-based and data-driven DT models to predict normal industrial behaviour and detect cyberattacks, while employing an exponentially weighted moving average based dynamic threshold to improve attack detection performance in Industry~4.0 manufacturing processes~\cite{Lin2023DT4I4-Secure:Security}. Along the same direction, a DT-enabled cyberattack detection framework integrates physics-based models, data-driven ML, and domain knowledge to distinguish cyberattacks from expected process anomalies during transient manufacturing operations under closed-loop control, demonstrating its effectiveness on a 3D printing system~\cite{Balta2024DigitalSystems}. Similarly, a DT-enabled ML framework for Industry~5.0 analyses DT-generated operational data for real-time detection of Distributed DoS (DDoS) attacks~\cite{Gaurav2023ALearning}.


Incident Response and Mitigation is addressed through a process-based DT security framework that integrates DT security simulations into Security Operations Centre (SOC) workflows to support cybersecurity monitoring and incident analysis~\cite{Dietz2020IntegratingCenter}. In the context of incident response, different studies investigated the use of DT to support cybersecurity incident response in CPS by analysing how different DT modalities can enhance incident response activities~\cite{Allison2023DigitalSystems, AkereleAbimbola2023TheInfrastructure}. Beyond attack detection, a DT combined with virtual actuator techniques enables both the detection and mitigation of deception attacks targeting control signals in cloud control systems, maintaining stable plant operation during cyberattacks~\cite{Akbarian2023DetectingTwins}. Extending this concept to complete ICS protection, a DT-based security framework integrates a digital-domain IDS with attack mitigation mechanisms to detect cyberattacks in a timely manner while preserving system stability during attacks~\cite{Akbarian2021AMitigation}.


Table~\ref{tab:dt_comparison} summarises the representative ICS cybersecurity DTs reviewed in this study. The DTs are classified according to their primary cybersecurity purposes and compared based on their implementation approaches, AI-enabled capabilities, and overall functionalities. 

\begin{table*}[t]
\caption{Comparison of Representative ICS Cybersecurity DTs Platforms}
\label{tab:dt_comparison}
\centering
\scriptsize
\begin{tabular}{
p{0.13\textwidth}
p{0.15\textwidth}
p{0.10\textwidth}
p{0.10\textwidth}
p{0.10\textwidth}
p{0.20\textwidth}
p{0.04\textwidth}
}

\toprule
\textbf{Category} &
\textbf{Digital Twin Platform} &
\textbf{Domain} &
\textbf{Implementation} &
\textbf{Primary Purpose} &
\textbf{Key Capabilities} &
\textbf{AI} \\
\midrule

\multirow{3}{*}{\parbox{2.5cm}{\raggedright Operational Monitoring \& Simulation}}

& IoT Motor DT
& Manufacturing
& Physical
& Condition Monitoring
& Real-time monitoring, predictive maintenance, visualisation
& $\times$ \\

& Cocoa Production DT
& Manufacturing
& Physical
& Process Monitoring
& Process simulation, PLC/HMI integration, engineering education
& $\times$ \\

& Cyber Twin
& Industry 4.0
& Virtual
& Virtual Engineering
& Semantic modelling, bidirectional communication, simulation
& $\times$ \\
\midrule

\multirow{7}{*}{\parbox{2cm}{\raggedright Security Testing \& Validation}}

& Industrial Network DT
& Generic ICS
& Virtual
& Cyber Range
& Attack experimentation, security validation
& $\times$ \\

& Filling Plant DT
& Manufacturing
& Virtual
& Attack Simulation
& Synthetic dataset generation, IDS evaluation
& \checkmark \\

& TaS
& IoT/CPS
& Virtual
& Software Validation
& Fault injection, what-if analysis
& $\times$ \\

& Fuzz Testing DT
& Generic ICS
& Virtual
& Vulnerability Assessment
& Automated fuzz testing, security verification
& $\times$ \\

& Smart Grid DT
& Smart Grid
& Virtual
& Penetration Testing
& Lifecycle security evaluation
& $\times$ \\

& Hybrid Smart Grid DT
& Smart Grid
& Hybrid
& Cybersecurity Evaluation
& Hybrid simulation, CPS emulation
& $\times$ \\

& Critical Infrastructure DT
& Critical Infrastructure
& Virtual
& Cybersecurity Analysis
& Network modelling, blockchain-enabled security
& $\times$ \\
\midrule

\multirow{5}{*}{\parbox{2.5cm}{\raggedright AI-based Detection \& Prediction}}

& Edge AI DT
& Generic ICS
& Virtual
& Attack Detection
& Offline ML training, edge deployment
& \checkmark \\

& Knowledge Graph DT
& Smart Farm
& Virtual
& Anomaly Detection
& Knowledge graph, ontology, ML
& \checkmark \\

& DT4I4-Secure
& Manufacturing
& Hybrid
& Cyberattack Detection
& Hybrid DT, predictive analytics
& \checkmark \\

& Physics-guided DT
& Manufacturing
& Hybrid
& Attack Detection
& Physics-informed modelling, ML
& \checkmark \\

& Industry 5.0 DT
& IIoT
& Virtual
& DDoS Detection
& ML-based cyberattack detection
& \checkmark \\
\midrule

\multirow{5}{*}{\parbox{2.5cm}{\raggedright Incident Response \& Mitigation}}

& SOC DT
& Generic ICS
& Virtual
& Incident Analysis
& SOC integration, security monitoring
& $\times$ \\

& DT Incident Response
& CPS
& Virtual
& Incident Response
& Response planning and analysis
& $\times$ \\

& DT Modalities
& CPS
& Virtual
& Incident Response
& Evaluation of DT-assisted response
& $\times$ \\

& Virtual Actuator DT
& Cloud Control Systems
& Hybrid
& Attack Mitigation
& Deception attack detection and mitigation
& $\times$ \\

& DT Security Framework
& Generic ICS
& Hybrid
& System Resilience
& IDS integration, attack mitigation
& $\times$ \\

\bottomrule
\end{tabular}

\begin{flushleft}
\footnotesize
\textit{Note:} Implementation refers to the primary deployment model of the digital twin: \emph{Physical} (connected to real industrial assets), \emph{Virtual} (software-based virtual representation), or \emph{Hybrid} (combining virtual models with physical hardware or HIL components). AI indicates whether AI or ML forms a core component of the DT platform.
\end{flushleft}
\end{table*}

\section{Datasets, Testbeds, and DTs: When and How to Use Them}

Datasets, testbeds, and DTs support different stages of AI-based ICS cybersecurity research. Datasets provide benchmark data for model training and evaluation, testbeds enable controlled experimentation under realistic industrial conditions, and DTs provide high-fidelity virtual replicas for continuous analysis and validation. As summarised in Table~\ref{tab:comparison_resources}, they differ in realism, data generation, deployment cost, and intended applications.

Datasets serve the purpose of modelling and benchmarking, while testbeds are used for executing attacks, data creation, and validation of the experiment prior to implementation. DTs further extend these capabilities by supporting real-time monitoring, predictive analysis, and continuous validation through virtual representations of physical systems. Consequently, the selection of an appropriate resource depends on the research objective. Datasets are most suitable for model development and benchmarking, testbeds are preferred for experimental validation and attack emulation, while DTs enable continuous evaluation and operational decision support. In practice, these resources should be viewed as complementary rather than competing technologies, forming an integrated research ecosystem that supports the complete lifecycle of AI-based ICS cybersecurity research. This complementary relationship forms the foundation of the roadmap presented in Section~\ref{sec:roadmap}.

\begin{table}
\caption{Comparison of Datasets, Testbeds, and DTs}
\label{tab:comparison_resources}
\centering
\scriptsize
\begin{tabular}{p{3cm}p{3cm}p{3cm}p{4cm}}
\toprule
\textbf{Aspect} & \textbf{Datasets} & \textbf{Testbeds} & \textbf{DTs} \\
\midrule
Primary purpose & Model development & Experimental validation & Continuous monitoring and optimisation \\

Role in AI lifecycle & Training \& benchmarking & Validation \& testing & Deployment \& continuous evaluation \\

Realism & Low--Medium & High & Very High \\
Physical equipment & Not required & Required (fully or partially) & Optional (connected to physical assets) \\

Data generation & Fixed (offline) & Dynamic & Continuous (real-time) \\

Cyberattack execution & No & Yes & Yes \\

Attack simulation & No & Yes & Yes (what-if analysis) \\

Model training & Excellent & Limited & Limited \\
Model validation & Limited & Excellent & Excellent \\
Real-time monitoring & No & Limited & Yes \\
Synthetic data generation & No & Limited & Yes \\
Deployment cost & Low & Medium--High & High \\
Scalability & High & Medium & High \\
Typical users & AI researchers & Cybersecurity researchers & Industry practitioners \\
\bottomrule
\end{tabular}
\end{table}

\section{Research Gaps and Open Challenges}

Despite significant advances in ICS testbeds for cybersecurity research, several challenges remain in achieving realistic, scalable, and reproducible experimental environments. The main challenge in testbed design is to balance between testbed fidelity and scalability. Physical testbeds have a very high level of fidelity, because of the presence of real industrial devices; however, they are costly, hard to set up and maintain. Simulation and virtualization technologies have high scalability but poor fidelity in terms of real behaviour of hardware, physical processes, communication delays and unpredicted situations. It becomes especially critical for CPS, since in this case the accurate modelling of processes and network environment is a key point.

The second difficulty is the diversity of the ICS testbeds. The industrial systems use a great number of heterogeneous devices and various vendors' equipment, whereas testbeds usually cover only some of them. In addition to that, the combination of legacy and modern technology in the industrial systems makes the testing even harder due to the insufficient capabilities of traditional IT security solutions for OT.

Operational constraints also introduce significant challenges. ICS applications have varying requirements for latency, sampling rates, reliability, and availability, where communication delays, packet loss, and network failures can affect system stability. Consequently, cybersecurity solutions should be evaluated under realistic operational conditions that consider both cyber and physical impacts.

Scalability is still a problem in the testbeds' design, as it is costly and complicated to build up the physical infrastructure. However, the scalable solutions such as virtualisation, HIL and simulation cannot replace physical components in the analysis of the CPSs. The future testbeds should have modular and extensible design to provide capability to connect new devices, networks and simulated processes.

The limited availability of realistic and diverse ICS cybersecurity datasets is another major research gap. AI-based anomaly detection approaches require datasets capturing normal operational behaviour, process dynamics, and diverse attack scenarios. However, collecting such data from real environments is challenging due to safety, confidentiality, and operational constraints. Moreover, effective security monitoring requires appropriate data collection and feature selection strategies, as excessive logging increases computational overhead while insufficient data may reduce detection effectiveness.

Another limitation associated with the comparison of ICS security solutions is that there is no standard approach to evaluating them. Indeed, each study uses its own dataset, attack scenarios and metrics, which prevents researchers from comparing the effectiveness and applicability of their approaches. Therefore, benchmarking frameworks should be developed for this purpose.

Finally, despite the promising results achieved in ICS security through the application of AI techniques, their deployment in OT environment remains problematic because many approaches rely on IT assumptions rather than the specifics of the OT environment, such as deterministic communication, process dependencies, safety constraints and computational limitations. Further research should pay more attention to the development of lightweight and explainable AI~\cite{Green2017PainsResearch, Smadi2021AChallenges}.

\section{Roadmap for Next-Generation ICS Dataset, Testbed and Digital Twins Development}
\label{sec:roadmap}.

Prior research has already developed frameworks and approaches related to creating ICS datasets, designing cyber ranges, constructing DTs, and benchmarking AI-based cybersecurity systems. Building upon the above-mentioned contributions, this section offers a synthesis of the literature into a roadmap outlining future work on dataset and testbed development.

ICSSIM provides a modular framework for constructing customised virtual and hybrid ICS testbeds~\cite{Dehlaghi-Ghadim2023ICSSIMTestbeds}, while Maynard \textit{et al.} introduced an automated framework for deploying virtual SCADA environments supporting multiple industrial communication protocols~\cite{Maynard2018AnNetworks}. For dataset development, a data-flow-based methodology was proposed to systematically generates attack scenarios from process control information, reducing manual effort and improving data consistency~\cite{Choi2022Dataflow-basedDevelopment}. More recently, DTs frameworks have been adopted to support real-time cybersecurity evaluation and AI-based IDS~\cite{Krishnaveni2024CyberDefender:Systems}, whereas AI-oriented frameworks have focused on addressing data imbalance, hyperparameter optimisation, and anomaly detection performance~\cite{Li2025STRTNet:Rebalancing}. Future ICS research environments should evolve towards scalable, realistic, and reproducible platforms that address the limitations of current datasets and testbeds. Key priorities include improving testbed fidelity, increasing device and protocol diversity, enabling efficient scalability, and supporting the generation of high-quality datasets suitable for AI-based security research.

These works offer practical advice for designing future experimental environments for cybersecurity research in ICS systems. They allow researchers to take advantage of these frameworks and thus ease the process of creating real-life datasets and testbeds, make the experiments more replicable, and create benchmarks for assessing the performance of AI-based solutions.

\begin{figure*}[htbp]
    \centering
    \includegraphics[width=0.9\textwidth]{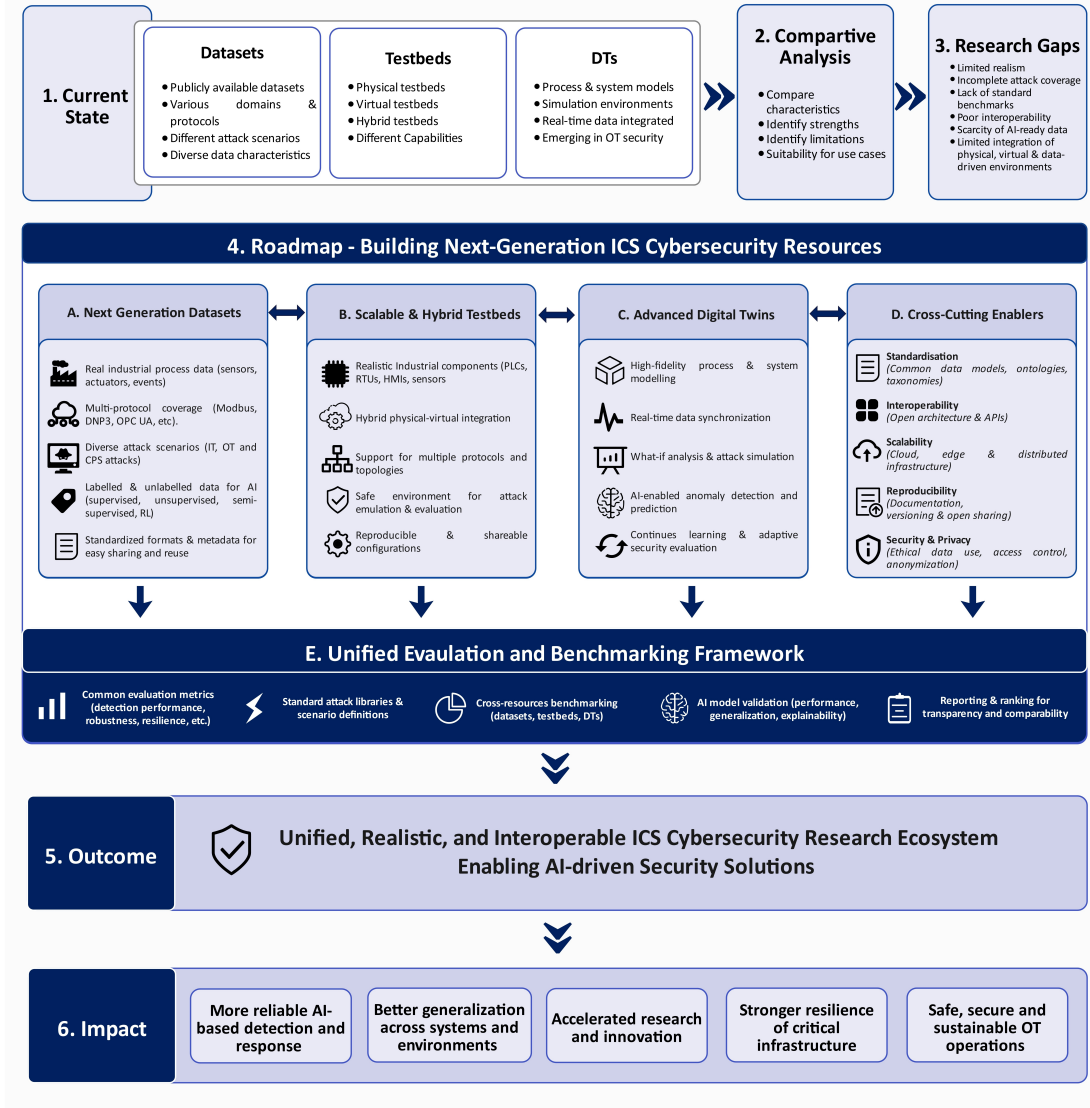}
    \caption{Roadmap of next-generation cybersecurity resources.}
    \label{fig:roadmap}
\end{figure*}

The roadmap proposed for the development of new-generation ICS cybersecurity resources is summarized in Figure~\ref{fig:roadmap}. The roadmap provides a systematic workflow to guide researchers from setting up research objectives and identifying appropriate development environments to designing industrial processes, creating cyberattack scenarios, collecting and validating quality datasets, and assessing AI-powered cybersecurity techniques. The roadmap also emphasizes the importance of considerations such as realism, scalability, reproducibility, variety of protocols, and validation, which are crucial for dataset, testbed, and digital twin creation.

\section{ Recommendations for Researchers and Practitioners}

The following recommendations could be used to develop and evaluate AI-based cybersecurity solutions for ICS on the basis of the conducted survey findings.

\begin{itemize}
\item Datasets selection must be determined by the aim of the research and not its popularity. Researcher has to pay attention to application domain, attack diversity, industrial communication protocols and characteristics of the data when selecting benchmark datasets.

\item Proposed solution must be evaluated in several different environments. Even though public datasets provide the possibility to benchmark the approach in a reproducible manner, the testbed and DT allow validating the approach in a realistic environment.

\item Introduce various cyber-physical attack cases. Attacks on ICS like FDIs, replay, DoS, MitM, malware, and attacks based on specific protocols should be included in future datasets and testbeds.

\item Provide a support of several industrial communication protocols (Modbus/TCP, DNP3, IEC~61850, OPC UA, BACnet, MQTT).

\item Provide high-quality datasets with good documentation, labels, ground truth and sufficient meta-information.

\item Promote open and standardised benchmark resources. Publicly available datasets, testbeds, and evaluation metrics will enable consistent performance assessment and accelerate the development of robust AI-based cybersecurity solutions.

\item Integrate DTs into the evaluation lifecycle to support safe experimentation, continuous validation, synthetic data generation, and testing before deployment in operational environments.
\end{itemize}

\begin{figure}[htbp]
    \centering
    \includegraphics[width=0.9\textwidth]{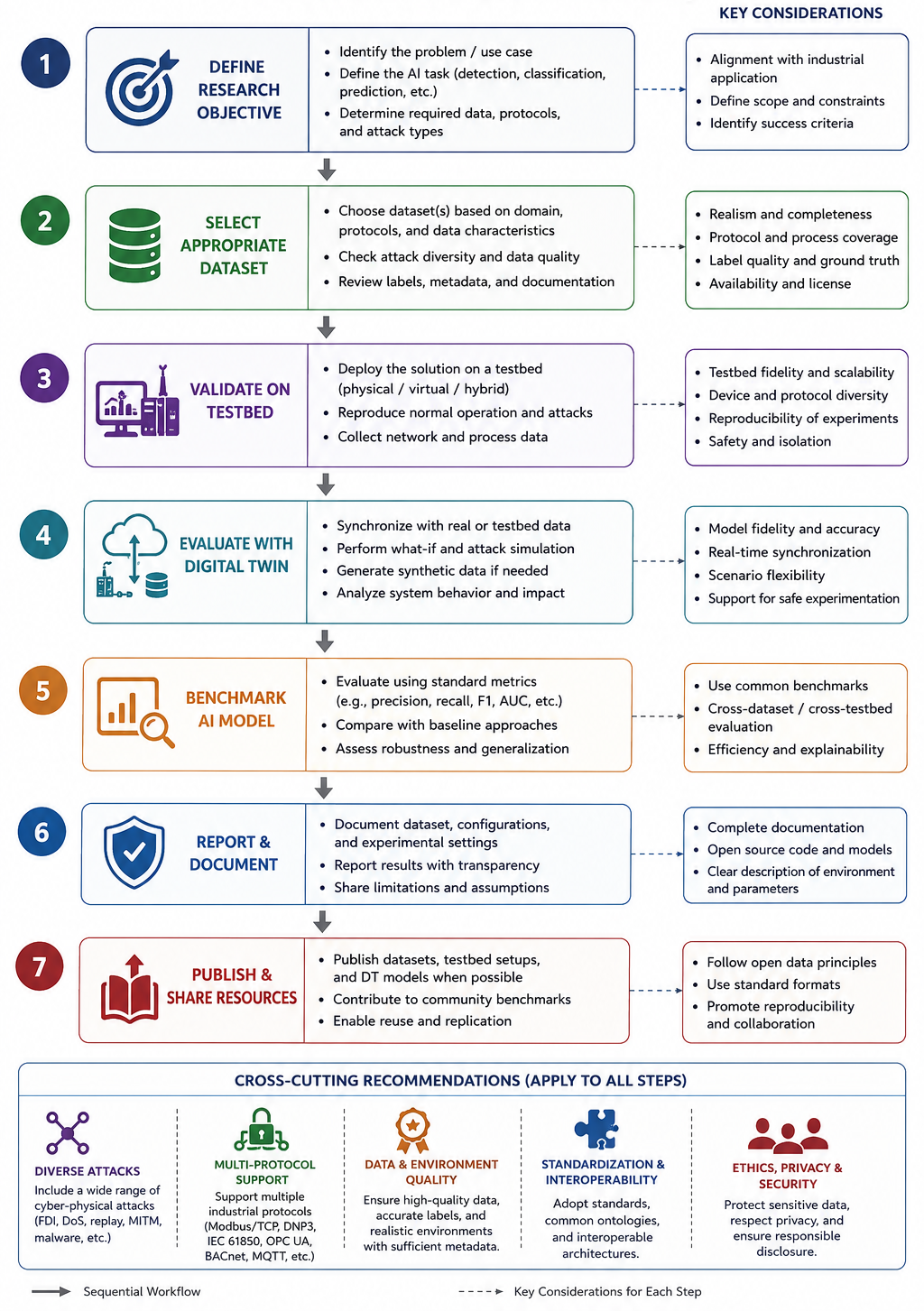}
    \caption{Recommended workflow for ICS Cybersecurity Research.}
    \label{fig:recommendation_workflow}
\end{figure}

Figure~\ref{fig:recommendation_workflow} shows an example of such workflow that represents how the major recommendations from this survey can be used to develop a research procedure. This workflow stresses on the need for conducting the experiment systematically by first defining the objective of the research, followed by selecting relevant cybersecurity resources and creating relevant attack scenarios. The other important thing is that validation of the AI model needs to be conducted in different environments, prior to benchmarking and deployment.

\section{Conclusion}

In this paper, we provide an extensive review of ICS cybersecurity datasets, testbeds, and DTs that can be used in AI security research. The selected resources were categorised depending on their use cases, data attributes, communication protocols, attack scenarios, and research purposes. A taxonomy was created to facilitate selection of suitable resources by indicating their benefits, drawbacks, and applicability for different cybersecurity tasks. Moreover, the critical issues and recommendations for future development of ICS cybersecurity resources were proposed.

Even though considerable advancements have been made regarding the availability of ICS cybersecurity resources, a number of important challenges still remain. Namely, these include inconsistent benchmarking approaches, lack of industrial process representation, coverage problems, as well as poor physical-virtual-data integration. The above-listed challenges need to be addressed further. Therefore, it is crucial to develop standardized and diverse datasets, scalable hybrid testbeds, AI-enabled experimental environments, and evaluation approaches.

\newpage
\bibliographystyle{plain}
\bibliography{references}

@misc{1998Laboratory,
    title = {{1998 DARPA Intrusion Detection Evaluation Dataset | MIT Lincoln Laboratory}},
    url = {https://www.ll.mit.edu/r-d/datasets/1998-darpa-intrusion-detection-evaluation-dataset},
    author = {MIT Lincoln Laboratory},
}

@article{Akbarian2021AMitigation,
    title = {{A Security Framework in Digital Twins for Cloud-based Industrial Control Systems: Intrusion Detection and Mitigation}},
    year = {2021},
    journal = {IEEE International Conference on Emerging Technologies and Factory Automation, ETFA},
    author = {Akbarian, Fatemeh and Tarneberg, William and Fitzgerald, Emma and Kihl, Maria},
    volume = {2021-September},
    publisher = {Institute of Electrical and Electronics Engineers Inc.},
    isbn = {9781728129891},
    doi = {10.1109/ETFA45728.2021.9613545},
    issn = {19460759}
}

@article{Beaver2013AnCommunications,
    title = {{An evaluation of machine learning methods to detect malicious SCADA communications}},
    year = {2013},
    journal = {Proceedings - 2013 12th International Conference on Machine Learning and Applications, ICMLA 2013},
    author = {Beaver, Justin M. and Borges-Hink, Raymond C. and Buckner, Mark A.},
    pages = {54--59},
    volume = {2},
    publisher = {IEEE Computer Society},
    doi = {10.1109/ICMLA.2013.105}
}

@article{Dehlaghi-GhadimAnomalySystems,
    title     = {{Anomaly Detection Dataset for Industrial Control Systems}},
    author    = {Dehlaghi-Ghadim, Alireza and Moghadam, Mahshid Helali and Balador, Ali and Hansson, Hans},
    journal   = {IEEE Access},
    volume    = {11},
    pages     = {107982--107996},
    year      = {2023},
    doi       = {10.1109/ACCESS.2023.3320928},
    url       = {https://www.kaggle.com/datasets/alirezadehlaghi/icssim},
    arxivId   = {2305.09678v1}
}

@misc{AutomationRevolution,
    title = {{Automation Trends Following Industrial Revolution}},
    url = {https://www.researchgate.net/publication/340332992_Automation_Trends_Following_Industrial_Revolution?channel=doi&linkId=5e84174da6fdcca789e596e0&showFulltext=true},
    author = {Muhammad Akbar Ramadhan and Natasya Zhafira and Sylvania Mulia FauziyyahUtari Trinita and Utari Trinita}
}

@misc{BATADAL,
    title  = {{BATADAL}},
    author = {Taormina, Riccardo and Galelli, Stefano and Tippenhauer, Nils O. and Salomons, Elad and Ostfeld, Avi and Eliades, Demetrios G. and A. R. C. and others},
    url    = {https://www.batadal.net/index.html}
}

@article{ElectraDataset,
    title        = {{On the Generation of Anomaly Detection Datasets in Industrial Control Systems}},
    author       = {Cervera, Carlos and others},
    journal      = {IEEE Access},
    volume       = {7},
    pages        = {177460--177473},
    year         = {2019},
    month        = dec,
    doi          = {10.1109/ACCESS.2019.2958284},
    url          = {http://perception.inf.um.es/ICS-datasets/}
}

@article{Adepu2019EPIC:Security,
    title = {{EPIC: An Electric Power Testbed for Research and Training in Cyber Physical Systems Security}},
    year = {2019},
    journal = {Lecture Notes in Computer Science (including subseries Lecture Notes in Artificial Intelligence and Lecture Notes in Bioinformatics)},
    author = {Adepu, Sridhar and Kandasamy, Nandha Kumar and Mathur, Aditya},
    pages = {37--52},
    volume = {11387 LNCS},
    publisher = {Springer, Cham},
    url = {https://link.springer.com/chapter/10.1007/978-3-030-12786-2_3},
    isbn = {978-3-030-12786-2},
    doi = {10.1007/978-3-030-12786-2{\_}3},
    issn = {1611-3349}
}

@inproceedings{ShinHAIDataset,
    title     = {{HAI 1.0: HIL-based Augmented ICS Security Dataset}},
    author    = {Shin, Hyeok-Ki and Lee, Woomyo and Yun, Jeong-Han and Kim, HyoungChun},
    booktitle = {13th USENIX Workshop on Cyber Security Experimentation and Test (CSET 2020)},
    year      = {2020},
    publisher = {USENIX Association},
    url       = {https://www.usenix.org/conference/cset20/presentation/shin},
    doi       = {10.5555/3485754.3485755}
}

@misc{IDSUNBb,
    title  = {{IDS 2017 | Datasets | Research | Canadian Institute for Cybersecurity | UNB}},
    author = {Sharafaldin, Iman and Lashkari, Arash Habibi and Ghorbani, Ali A.},
    url    = {https://www.unb.ca/cic/datasets/ids-2017.html}
}

@misc{IMPACTDatasets,
    title  = {{IMPACT - ICS Cyber Attack Gas Pipeline Datasets}},
    author = {{IMPACT Cyber Trust}},
    url    = {https://www.impactcybertrust.org/dataset_view?idDataset=1322}
}

@article{Panwar2019ImplementationWEKA,
    title = {{Implementation of machine learning algorithms on cicids-2017 dataset for intrusion detection using WEKA}},
    year = {2019},
    journal = {International Journal of Recent Technology and Engineering},
    author = {Panwar, Shailesh Singh and Negi, Pritam Singh and Panwar, Lokesh Singh and Raiwani, Y. P.},
    number = {3},
    month = {9},
    pages = {2195--2207},
    volume = {8},
    publisher = {Blue Eyes Intelligence Engineering and Sciences Publication},
    doi = {10.35940/IJRTE.C4587.098319},
    issn = {22773878}
}

@misc{BACnetInternationalIntroductionBACnet,
    title        = {{Introduction to BACnet}},
    author       = {{BACnet International}},
    year         = {2014},
    publisher    = {BACnet International},
    note         = {For Building Owners and Engineers, Version 1.0},
    url          = {https://www.ccontrols.com/pdf/BACnetIntroduction.pdf}
}

@misc{ITrustITrust,
    title  = {{iTrust Labs - SWaT - iTrust}},
    author = {{iTrust Centre for Research in Cyber Security}},
    url    = {https://itrust.sutd.edu.sg/itrust-labs-home/itrust-labs_swat/}
}

@misc{KDDData,
    title  = {{KDD Cup 1999 Data}},
    author = {Stolfo, Salvatore J. and Fan, Wei and Lee, Wenke and Prodromidis, Andreas and Chan, Philip K.},
    url    = {https://kdd.ics.uci.edu/databases/kddcup99/kddcup99.html}
}

@misc{NSL-KDDUNB,
    title  = {{NSL-KDD | Datasets | Research | Canadian Institute for Cybersecurity | UNB}},
    author = {Tavallaee, Mohammad and Bagheri, Ebrahim and Lu, Wei and Ghorbani, Ali A.},
    url    = {https://www.unb.ca/cic/datasets/nsl.html}
}

@article{Gomez2019OnSystems,
    title = {{On the Generation of Anomaly Detection Datasets in Industrial Control Systems}},
    year = {2019},
    journal = {IEEE Access},
    author = {G{\'{o}}mez, Ángel Luis Perales and Maim{\'{o}}, Lorenzo Fernández and Celdr{\'{a}}n, Alberto Huertas and Clemente, F. lix J.Garc a. and Sarmiento, Cristian Cadenas and Masa, Carlos Javier Del Canto and Nistal, Rub n.M.ndez},
    pages = {177460--177473},
    volume = {7},
    publisher = {Institute of Electrical and Electronics Engineers Inc.},
    doi = {10.1109/ACCESS.2019.2958284},
    issn = {21693536}
}

@misc{PowerState,
    title        = {{Power System Attack Datasets - Mississippi State University and Oak Ridge National Laboratory}},
    author       = {{Mississippi State University and Oak Ridge National Laboratory}},
    year         = {2014},
    month        = apr,
    url          = {https://ece.uah.edu/~thm0009/icsdatasets/PowerSystem_Dataset_README.pdf}
}

@inproceedings{LippmannResultsEvaluation,
    title     = {{Results of the DARPA 1998 Offline Intrusion Detection Evaluation}},
    author    = {Lippmann, Richard P. and Cunningham, Robert K. and Fried, David J. and Graf, Isaac and Kendall, Kris R. and Webster, Seth E. and Zissman, Marc A.},
    booktitle = {Recent Advances in Intrusion Detection (RAID 1999)},
    year      = {1999},
    pages     = {1--29}
}

@misc{SWaTDetection,
    title  = {{SWaT Testbed - Kaspersky Machine Learning for Anomaly Detection}},
    author = {{Kaspersky}},
    url    = {https://mlad.kaspersky.com/swat-testbed/}
}

@misc{TheResearch,
    title  = {{The UNSW-NB15 Dataset | UNSW Research}},
    author = {Moustafa, Nour and Slay, Jill},
    url    = {https://research.unsw.edu.au/projects/unsw-nb15-dataset}
}

@misc{ToN_IoTDataPort,
    title  = {{ToN{\_}IoT datasets | IEEE DataPort}},
    author = {Moustafa, Nour},
    url    = {https://ieee-dataport.org/documents/toniot-datasets}
}

@article{Booij2022ToN_IoT:Sets,
    title = {{ToN{\_}IoT: The Role of Heterogeneity and the Need for Standardization of Features and Attack Types in IoT Network Intrusion Data Sets}},
    year = {2022},
    journal = {IEEE Internet of Things Journal},
    author = {Booij, Tim M. and Chiscop, Irina and Meeuwissen, Erik and Moustafa, Nour and Hartog, Frank T.H.Den},
    number = {1},
    month = {1},
    pages = {485--496},
    volume = {9},
    publisher = {Institute of Electrical and Electronics Engineers Inc.},
    doi = {10.1109/JIOT.2021.3085194},
    issn = {23274662}
}

@article{Moustafa2015UNSW-NB15:Set,
    title = {{UNSW-NB15: A comprehensive data set for network intrusion detection systems (UNSW-NB15 network data set)}},
    year = {2015},
    journal = {2015 Military Communications and Information Systems Conference, MilCIS 2015 - Proceedings},
    author = {Moustafa, Nour and Slay, Jill},
    month = {12},
    publisher = {Institute of Electrical and Electronics Engineers Inc.},
    isbn = {9781467370080},
    doi = {10.1109/MILCIS.2015.7348942}
}

@misc{WUSTL-IIOT-2018Research,
   author = {Washington University in St. Louis},
   year = {2018},
    title = {{WUSTL-IIOT-2018 Dataset for ICS (SCADA) Cybersecurity Research}},
    url = {https://www.cse.wustl.edu/~jain/iiot/index.html}
}

@misc{WUSTL-IIOT-2018DataPort,
    title  = {{WUSTL-IIOT-2018 | IEEE DataPort}},
    author = {Teixeira, Marcio Andrey and Zolanvari, Maede and Jain, Raj},
    url    = {https://ieee-dataport.org/open-access/wustl-iiot-2018}
}

@misc{NetworkAttacks,
    title  = {{Network Traffic Dataset IEC-61850 Cyber Attacks}},
    author = {{Johnatan de Oliveira}},
    url    = {https://www.kaggle.com/datasets/johnatandeoliveira/network-traffic-dataset-iec-61850-cyber-attacks}
}

@misc{DNP3,
doi = {10.21227/s7h0-b081},
url = {https://dx.doi.org/10.21227/s7h0-b081},
author = {Panagiotis Radoglou-Grammatikis and Vasiliki Kelli and Thomas Lagkas and Vasileios Argyriou and Panagiotis Sarigiannidis},
publisher = {IEEE Dataport},
title = {DNP3 Intrusion Detection Dataset},
year = {2022} }

@misc{BACnetDataset,
    title = {{BACnet network dataset}},
    url = {https://www.kaggle.com/datasets/78eb45aeaac481853135d90738672111e46fe2a4ce653573e8856fc920f3da68},
    author = {Seyed Amirhossein Moosavi and Mojtaba Asgari and Seyed Reza Kamel}
}

@article{Ferrag2022Edge-IIoTset:Learning,
    title = {{Edge-IIoTset: A New Comprehensive Realistic Cyber Security Dataset of IoT and IIoT Applications for Centralized and Federated Learning}},
    year = {2022},
    journal = {IEEE Access},
    author = {Ferrag, Mohamed Amine and Friha, Othmane and Hamouda, Djallel and Maglaras, Leandros and Janicke, Helge},
    pages = {40281--40306},
    volume = {10},
    publisher = {Institute of Electrical and Electronics Engineers Inc.},
    doi = {10.1109/ACCESS.2022.3165809},
    issn = {21693536}
}

@article{Al-Hawawreh2022X-IIoTID:Things,
    title = {{X-IIoTID: A Connectivity-Agnostic and Device-Agnostic Intrusion Data Set for Industrial Internet of Things}},
    year = {2022},
    journal = {IEEE Internet of Things Journal},
    author = {Al-Hawawreh, Muna and Sitnikova, Elena and Aboutorab, Neda},
    number = {5},
    month = {3},
    pages = {3962--3977},
    volume = {9},
    publisher = {Institute of Electrical and Electronics Engineers Inc.},
    doi = {10.1109/JIOT.2021.3102056},
    issn = {23274662}
}

@misc{MQTT-IoT-IDS2020:DataPort,
    title  = {{MQTT-IoT-IDS2020: MQTT Internet of Things Intrusion Detection Dataset | IEEE DataPort}},
    author = {Hindy, Hanan and Tachtatzis, Christos and Atkinson, Robert C. and Bayne, Ethan and Bellekens, Xavier J. A.},
    url    = {https://ieee-dataport.org/open-access/mqtt-iot-ids2020-mqtt-internet-things-intrusion-detection-dataset}
}

@article{Neto2023CICIoT2023:Environment,
    title = {{CICIoT2023: A Real-Time Dataset and Benchmark for Large-Scale Attacks in IoT Environment}},
    year = {2023},
    journal = {Sensors 2023, Vol. 23, Page 5941},
    author = {Neto, Euclides Carlos Pinto and Dadkhah, Sajjad and Ferreira, Raphael and Zohourian, Alireza and Lu, Rongxing and Ghorbani, Ali A.},
    number = {13},
    month = {6},
    pages = {5941},
    volume = {23},
    publisher = {Multidisciplinary Digital Publishing Institute},
    url = {https://www.mdpi.com/1424-8220/23/13/5941/htm https://www.mdpi.com/1424-8220/23/13/5941},
    doi = {10.3390/S23135941},
    issn = {1424-8220},
    pmid = {37447792}
}

@article{Alanazi2025ICS-LTU2022:Vulnerabilities,
    title = {{ICS-LTU2022: A dataset for ICS vulnerabilities}},
    year = {2025},
    journal = {Computers {\&} Security},
    author = {Alanazi, Manar and Mahmood, Abdun and Chowdhury, Mohammad Jabed Morshed},
    month = {1},
    pages = {104143},
    volume = {148},
    publisher = {Elsevier Advanced Technology},
    doi = {10.1016/J.COSE.2024.104143},
    issn = {0167-4048}
}

@article{Yusof2025SignalsRange,
    title = {{Signals and Symptoms: ICS Attack Dataset From Railway Cyber Range}},
    year = {2025},
    journal = {arXiv.org},
    author = {Yusof, Anis and Liu, Yuancheng and Kang, N. and Seah, C. and Liang, Zhenkai and Chang, Ee-Chien},
    doi = {10.48550/ARXIV.2507.01768}
}

@article{Gaggero2024IndustrialEnvironments,
    title = {{Industrial Control System-Anomaly Detection Dataset (ICS-ADD) for Cyber-Physical Security Monitoring in Smart Industry Environments}},
    year = {2024},
    journal = {IEEE Access},
    author = {Gaggero, Giovanni Battista and Armellin, Alessandro and Portomauro, Giancarlo and Marchese, Mario},
    pages = {64140--64149},
    volume = {12},
    publisher = {Institute of Electrical and Electronics Engineers Inc.},
    doi = {10.1109/ACCESS.2024.3395991},
    issn = {21693536}
}

@article{Cabaj2025Cyber4OTSystems,
    title = {{Cyber4OT dataset: Network traces for cyber-security vulnerability evaluation in industrial control systems}},
    year = {2025},
    journal = {SoftwareX},
    author = {Cabaj, Krzysztof and Plamowski, Sebastian and Chaber, Patryk and {\L}awry{\'{n}}czuk, Maciej and Marusak, Piotr and Nebeluk, Robert and Wojtulewicz, Andrzej and Zarzycki, Krzysztof},
    month = {9},
    pages = {102196},
    volume = {31},
    publisher = {Elsevier},
    doi = {10.1016/J.SOFTX.2025.102196},
    issn = {2352-7110}
}

@article{Abukeshek2026STSRS:Systems,
    title = {{STSRS: A New Dataset for Simulating Security Threats in Smart Railway Systems}},
    year = {2026},
    journal = {International Conference on Information Systems Security and Privacy},
    author = {Abukeshek, Mays and Al-mhiqani, Mohammed and Parkinson, Simon},
    month = {3},
    pages = {292--299},
    publisher = {INSTICC},
    doi = {10.5220/0014572400004061}
}

@article{Dehlaghi-Ghadim2023ICSSIMTestbeds,
    title = {{ICSSIM — A framework for building industrial control systems security testbeds}},
    year = {2023},
    journal = {Computers in Industry},
    author = {Dehlaghi-Ghadim, Alireza and Balador, Ali and Moghadam, Mahshid Helali and Hansson, Hans and Conti, Mauro},
    month = {6},
    pages = {103906},
    volume = {148},
    publisher = {Elsevier},
    doi = {10.1016/J.COMPIND.2023.103906},
    issn = {0166-3615}
}

@article{Maynard2018AnNetworks,
    title     = {{An Open Framework for Deploying Experimental SCADA Testbed Networks}},
    author    = {Maynard, Peter and McLaughlin, Kieran and Sezer, Sakir},
    journal   = {International Workshop on Cyber-Physical Security (CPS 2018)},
    year      = {2018},
    publisher = {BCS Learning \& Development},
    doi       = {10.14236/EWIC/ICS2018.11}
}

@article{Choi2022Dataflow-basedDevelopment,
    title = {{Dataflow-based Control Process Identification for ICS Dataset Development}},
    year = {2022},
    journal = {ACM International Conference Proceeding Series},
    author = {Choi, Seungoh and Shin, Hyeok Ki and Lee, Woomyo and Yun, Jeong Han and Min, Byung Gil},
    month = {8},
    pages = {54--58},
    publisher = {Association for Computing Machinery},
    url = {https://dl-acm-org.kfupm.idm.oclc.org/doi/10.1145/3546096.3546113},
    isbn = {9781450396844},
    doi = {10.1145/3546096.3546113}
}

@article{Krishnaveni2024CyberDefender:Systems,
    title = {{CyberDefender: an integrated intelligent defense framework for digital-twin-based industrial cyber-physical systems}},
    year = {2024},
    journal = {Cluster Computing},
    author = {Krishnaveni, S. and Chen, Thomas M. and Sathiyanarayanan, Mithileysh and Amutha, B.},
    number = {6},
    month = {9},
    pages = {7273--7306},
    volume = {27},
    publisher = {Springer},
    url = {https://link.springer.com/article/10.1007/s10586-024-04320-x},
    doi = {10.1007/S10586-024-04320-X/METRICS},
    issn = {15737543}
}

@article{Li2025STRTNet:Rebalancing,
    title = {{STRTNet: An ICS Anomaly Detection Model With Hyperparameter Selection and Data Rebalancing}},
    year = {2025},
    journal = {IEEE Access},
    author = {Li, Tingting and Ma, Lulu and Li, Weili and Meng, Xin},
    pages = {205774--205790},
    volume = {13},
    publisher = {Institute of Electrical and Electronics Engineers Inc.},
    doi = {10.1109/ACCESS.2025.3636897},
    issn = {21693536}
}

@article{Conti2021AResearch,
    title = {{A Survey on Industrial Control System Testbeds and Datasets for Security Research}},
    year = {2021},
    journal = {IEEE Communications Surveys and Tutorials},
    author = {Conti, Mauro and Donadel, Denis and Turrin, Federico},
    number = {4},
    pages = {2248--2294},
    volume = {23},
    publisher = {Institute of Electrical and Electronics Engineers Inc.},
    doi = {10.1109/COMST.2021.3094360},
    issn = {1553877X},
    arxivId = {2102.05631}
}

@article{Holm2015ATestbeds,
    title = {{A Survey of Industrial Control System Testbeds}},
    year = {2015},
    journal = {Nordic Conference on Secure IT Systems},
    author = {Holm, Hannes and Karresand, Martin and Vidstr{\"{o}}m, Arne and Westring, Erik},
    pages = {11--26},
    volume = {9417},
    publisher = {Springer Verlag},
    isbn = {9783319265018},
    doi = {10.1007/978-3-319-26502-5{\_}2},
    issn = {16113349}
}

@article{Geng2019ATestbeds,
    title = {{A survey of industrial control system testbeds}},
    year = {2019},
    journal = {IOP Conference Series: Materials Science and Engineering},
    author = {Geng, Yangyang and Wang, Yi and Liu, Wenwen and Wei, Qiang and Liu, Ke and Wu, Haolan},
    number = {4},
    month = {8},
    volume = {569},
    publisher = {Institute of Physics Publishing},
    doi = {10.1088/1757-899X/569/4/042030},
    issn = {1757899X}
}

@article{Zhang2024AResearch,
    title     = {{A Survey on Industrial Internet of Things (IIoT) Testbeds for Connectivity Research}},
    year      = {2024},
    author    = {Zhang, Tianyu and Xue, Chuanyu and Wang, Jiachen and Yun, Zelin and Lin, Natong and Han, Song},
    journal   = {arXiv preprint arXiv:2404.17485},
    month     = {4},
    url       = {https://arxiv.org/abs/2404.17485v2},
    arxivId   = {2404.17485}
}

@article{El-HajjTaruItapeltoTeklitGebremariam2024SystematicApplications,
    title = {{Systematic literature review: Digital twins' role in enhancing security for Industry 4.0 applications}},
    year = {2024},
    journal = {Security and Privacy},
    author = {El-Hajj Taru It{\"{a}}pelto Teklit Gebremariam, Mohammed and Mohammed El-Hajj, Correspondence},
    number = {5},
    month = {9},
    pages = {e396},
    volume = {7},
    publisher = {John Wiley {\&} Sons, Ltd},
    url = {https://onlinelibrary.wiley.com/doi/full/10.1002/spy2.396 https://onlinelibrary.wiley.com/doi/abs/10.1002/spy2.396 https://onlinelibrary.wiley.com/doi/10.1002/spy2.396},
    doi = {10.1002/SPY2.396},
    issn = {2475-6725}
}

@article{Mendonca2022DigitalChallenges,
    title = {{Digital Twin Applications: A Survey of Recent Advances and Challenges}},
    year = {2022},
    journal = {Processes 2022, Vol. 10, Page 744},
    author = {Mendon{\c{c}}a, Rafael da Silva and Lins, Sidney de Oliveira and de Bessa, Iury Valente and de Carvalho Ayres, Florindo Antônio and de Medeiros, Renan Landau Paiva and de Lucena, Vicente Ferreira},
    number = {4},
    month = {4},
    pages = {744},
    volume = {10},
    publisher = {Multidisciplinary Digital Publishing Institute},
    url = {https://www.mdpi.com/2227-9717/10/4/744/htm https://www.mdpi.com/2227-9717/10/4/744},
    doi = {10.3390/PR10040744},
    issn = {2227-9717}
}

@article{Mun2025ARequirements,
    title = {{A Comprehensive Survey on Digital Twin: Focusing on Security Threats and Requirements}},
    year = {2025},
    journal = {IEEE Access},
    author = {Mun, Hyeran and Han, Kyusuk and Damiani, Ernesto and Yeun, Hyun Ku and Kim, Tae Yeon and Martino, Luigi and Yeun, Chan Yeob},
    pages = {73362--73390},
    volume = {13},
    publisher = {Institute of Electrical and Electronics Engineers Inc.},
    doi = {10.1109/ACCESS.2025.3563621},
    issn = {21693536}
}

@article{Termanini2026IntrusionRubric,
    title = {{Intrusion Detection Datasets for IIoT and ICS: A Taxonomic Review with a Decision-Aid Scoring Rubric}},
    year = {2026},
    journal = {Sensors 2026, Vol. 26, Page 4099},
    author = {Termanini, Ayman and Bourdoucen, Hadj and Al-Abri, Dawood and Maashri, Ahmed Al},
    number = {13},
    month = {6},
    pages = {4099},
    volume = {26},
    publisher = {Multidisciplinary Digital Publishing Institute},
    url = {https://www.mdpi.com/1424-8220/26/13/4099/htm https://www.mdpi.com/1424-8220/26/13/4099},
    doi = {10.3390/S26134099},
    issn = {1424-8220}
}

@article{Shin2021TwoTestbed,
    title = {{Two ICS Security Datasets and Anomaly Detection Contest on the HIL-based Augmented ICS Testbed}},
    year = {2021},
    journal = {ACM International Conference Proceeding Series},
    author = {Shin, Hyeok Ki and Lee, Woomyo and Yun, Jeong Han and Min, Byung Gil},
    month = {9},
    pages = {36--40},
    publisher = {Association for Computing Machinery},
    url = {https://dl.acm.org/doi/10.1145/3474718.3474719},
    isbn = {9781450390651},
    doi = {10.1145/3474718.3474719}
}

@article{Sauer2019LICSTERResearch,
    title     = {{LICSTER -- A Low-cost ICS Security Testbed for Education and Research}},
    author    = {Sauer, Felix and Niedermaier, Matthias and Kie{\ss}ling, Susanne and Merli, Dominik},
    journal   = {Electronic Workshops in Computing (eWiC)},
    year      = {2019},
    publisher = {BCS Learning \& Development},
    doi       = {10.14236/EWIC/ICSCSR19.1},
    arxivId   = {1910.00303}
}

@article{Ekisa2022VICSORT-ATestbed,
    title = {{VICSORT-A Virtualised ICS Open-source Research Testbed}},
    year = {2022},
    journal = {2022 Cyber Research Conference - Ireland, Cyber-RCI 2022},
    author = {Ekisa, Conrad and Briain, Diarmuid O. and Kavanagh, Yvonne},
    publisher = {Institute of Electrical and Electronics Engineers Inc.},
    isbn = {9781665474221},
    doi = {10.1109/CYBER-RCI55324.2022.10032670}
}

@article{Sicard2022AnResearch,
    title = {{An Industrial Control System Physical Testbed for Naval Defense Cybersecurity Research}},
    year = {2022},
    journal = {Proceedings - 7th IEEE European Symposium on Security and Privacy Workshops, Euro S and PW 2022},
    author = {Sicard, Franck and Hotellier, Estelle and Francq, Julien},
    pages = {413--422},
    publisher = {Institute of Electrical and Electronics Engineers Inc.},
    isbn = {9781665495608},
    doi = {10.1109/EUROSPW55150.2022.00049}
}

@article{Karch2022CrossTest:Evaluations,
    title = {{CrossTest: a cross-domain physical testbed environment for cybersecurity performance evaluations}},
    year = {2022},
    journal = {IEEE International Conference on Emerging Technologies and Factory Automation, ETFA},
    author = {Karch, Markus and Rosch, Dennis and Andre, Kummerow and Meshram, Ankush and Haas, Christian and Nicolai, Steffen},
    volume = {2022-September},
    publisher = {Institute of Electrical and Electronics Engineers Inc.},
    isbn = {9781665499965},
    doi = {10.1109/ETFA52439.2022.9921672},
    issn = {19460759}
}

@article{Garcia2025SPHEREExperimentation,
    title = {{SPHERE CPS Enclave: A Reconfigurable Testbed for Industrial Control System Security Experimentation}},
    year = {2025},
    journal = {Proceedings of the ACM/IEEE 16th International Conference on Cyber-Physical Systems, ICCPS 2025, held as part of the CPS-IoT Week 2025},
    author = {Garcia, Luis and Mirkovic, Jelena and Balenson, David and Kline, Erik and Pradkin, Yuri and Choffnes, David and Dubois, Daniel and Benzel, Terry and Ravi, Srivatsan and Barnes, Joseph and Lawler, Geoff and Tran, Chris and Regalado, Alba},
    month = {5},
    volume = {1},
    publisher = {Association for Computing Machinery, Inc},
    url = {https://dl-acm-org.kfupm.idm.oclc.org/doi/10.1145/3716550.3725158},
    isbn = {9798400714986},
    doi = {10.1145/3716550.3725158}
}

@article{Chhokra2025HIL-RESIST:Systems,
    title = {{HIL-RESIST: Hardware-In-the-Loop testbed for RESilience evaluation against Integrated System Threats in power systems}},
    year = {2025},
    journal = {MITOTI 2025 - Proceedings of the International Workshop on Middleware for IT/OT Integration, Part of Middleware Conference 2025},
    author = {Chhokra, Ajay Dev and Shekhar, Shashank and Bhela, Siddharth and Muenz, Ulrich},
    month = {12},
    pages = {5--8},
    volume = {1},
    publisher = {Association for Computing Machinery, Inc},
    url = {https://dl-acm-org.kfupm.idm.oclc.org/doi/10.1145/3774900.3776634},
    isbn = {9798400723032},
    doi = {10.1145/3774900.3776634}
}

@article{Raj2025AutonomousNetworks,
    title = {{Autonomous Cybersecurity Testbed for Operational Technology Networks}},
    year = {2025},
    journal = {Proceedings of the 2025 Workshop on Design Automation for CPS and IoT, DESTION 2025, 2025 Cyber-Physical Systems and Internet-of-Things Week, CPS-IoT Week 2025 Workshops},
    author = {Raj, Akhilesh and Das, Sanjana and Vardhan, Harsh and Neema, Himanshu and Chhokra, Ajay and Balasubramanian, Daniel},
    month = {6},
    publisher = {Association for Computing Machinery, Inc},
    url = {https://dl-acm-org.kfupm.idm.oclc.org/doi/10.1145/3722573.3727830},
    isbn = {9798400716041},
    doi = {10.1145/3722573.3727830}
}

@article{Rahmani2025AOT-Networks,
    title = {{A Testbed for Cyber Attack Emulation and AI-Driven Anomaly Detection in Industrial IoTand OT-Networks}},
    year = {2025},
    journal = {Proceedings of the IEEE International Conference on Intelligent Data Acquisition and Advanced Computing Systems: Technology and Applications, IDAACS},
    author = {Rahmani, Jaafer and Detken, Kai Oliver and Sikora, Axel},
    pages = {255--260},
    publisher = {Institute of Electrical and Electronics Engineers Inc.},
    isbn = {9798331580452},
    doi = {10.1109/IDAACS68557.2025.11322367},
    issn = {27704254}
}

@article{Ashok2021ASystems,
    title = {{A High-Fidelity Cyber-Physical Testbed-Based Benchmarking Dataset for Testing Operational Technology Specific Intrusion Detection Systems}},
    year = {2021},
    journal = {2021 IEEE Virtual IEEE International Symposium on Technologies for Homeland Security, HST 2021},
    author = {Ashok, Aditya and Edgar, Thomas},
    publisher = {Institute of Electrical and Electronics Engineers Inc.},
    isbn = {9781665441520},
    doi = {10.1109/HST53381.2021.9619851}
}

@article{Calder2023SWaP:Research,
    title = {{SWaP: A Water Process Testbed for ICS Security Research}},
    year = {2023},
    journal = {2023 IEEE International Conference on Omni-Layer Intelligent Systems, COINS 2023},
    author = {Calder, Matthew and Ahmed, Mujeeb and Nagaraja, Shishir},
    publisher = {Institute of Electrical and Electronics Engineers Inc.},
    isbn = {9798350346473},
    doi = {10.1109/COINS57856.2023.10189315}
}

@article{Khan2020LightweightSystems,
    title = {{Lightweight Testbed for Cybersecurity Experiments in SCADA-based Systems}},
    year = {2020},
    journal = {2020 International Conference on Computing and Information Technology, ICCIT 2020},
    author = {Khan, Mohsin and Rehman, Osama and Rahman, Ibrahim M.H. and Ali, Saqib},
    month = {9},
    publisher = {Institute of Electrical and Electronics Engineers Inc.},
    isbn = {9781728126807},
    doi = {10.1109/ICCIT-144147971.2020.9213791}
}

@article{Adhikari2017WAMSMining,
    title = {{WAMS Cyber-Physical Test Bed for Power System, Cybersecurity Study, and Data Mining}},
    year = {2017},
    journal = {IEEE Transactions on Smart Grid},
    author = {Adhikari, Uttam and Morris, Thomas and Pan, Shengyi},
    number = {6},
    month = {11},
    pages = {2744--2753},
    volume = {8},
    publisher = {Institute of Electrical and Electronics Engineers Inc.},
    doi = {10.1109/TSG.2016.2537210},
    issn = {19493053}
}

@article{Lo2025TRIST:Detection,
    title = {{TRIST: Towards a Container-Based ICS Testbed for Cyber Threat Simulation and Anomaly Detection}},
    year = {2025},
    journal = {Springer Proceedings in Complexity},
    author = {Lo, Carol and Christie, Jack and Win, Thu Yein and Rezaeifar, Zeinab and Khan, Zaheer and Legg, Phil},
    pages = {225--239},
    publisher = {Springer Science and Business Media B.V.},
    url = {https://link.springer.com/chapter/10.1007/978-981-96-0401-2_13},
    isbn = {9789819604005},
    doi = {10.1007/978-981-96-0401-2{\_}13/SAVE-RESEARCH},
    issn = {22138692}
}

@article{Sutterfield2024Real-TimeSystems,
    title = {{Real-Time Simulation and Workforce Development with TROY: Testbed for Resilient Operational Systems}},
    year = {2024},
    journal = {2024 IEEE Design Methodologies Conference, DMC 2024},
    author = {Sutterfield, Gideon and Schmidt, Henry and Farnell, Chris},
    publisher = {Institute of Electrical and Electronics Engineers Inc.},
    isbn = {9798350355864},
    doi = {10.1109/DMC62632.2024.10812136}
}

@article{Xie2018VTET:Research,
    title = {{VTET: A Virtual Industrial Control System Testbed for Cyber Security Research}},
    year = {2018},
    journal = {2018 3rd International Conference on Security of Smart Cities, Industrial Control System and Communications, SSIC 2018 - Proceedings},
    author = {Xie, Yaobin and Wang, Wei and Wang, Faren and Chang, Rui},
    month = {12},
    publisher = {Institute of Electrical and Electronics Engineers Inc.},
    isbn = {9781538681879},
    doi = {10.1109/SSIC.2018.8556732}
}

@article{Mekala2022IndustrialForensics,
    title = {{Industrial Internet of Things (IIoT): Testbed and Datasets for CyberSecurity and Digital Forensics}},
    year = {2022},
    journal = {2022 International Conference on Smart Generation Computing, Communication and Networking, SMART GENCON 2022},
    author = {Mekala, Sri Harsha and Baig, Zubair and Anwar, Adnan},
    publisher = {Institute of Electrical and Electronics Engineers Inc.},
    isbn = {9781665454995},
    doi = {10.1109/SMARTGENCON56628.2022.10083609}
}

@article{Jarmakiewicz2015DevelopmentInfrastructure,
    title = {{Development of cyber security testbed for critical infrastructure}},
    year = {2015},
    journal = {2015 International Conference on Military Communications and Information Systems, ICMCIS 2015},
    author = {Jacek Jarmakiewicz and Krzysztof Ma{\"s}lanka and Krzysztof Parobczak},
    month = {7},
    publisher = {Institute of Electrical and Electronics Engineers Inc.},
    isbn = {9788393484812},
    doi = {10.1109/ICMCIS.2015.7158687}
}

@article{Reaves2012AnResearch,
    title = {{An open virtual testbed for industrial control system security research}},
    year = {2012},
    journal = {International Journal of Information Security},
    author = {Reaves, Bradley and Morris, Thomas},
    number = {4},
    month = {8},
    pages = {215--229},
    volume = {11},
    publisher = {Springer},
    url = {https://link.springer.com/article/10.1007/s10207-012-0164-7},
    doi = {10.1007/S10207-012-0164-7/METRICS},
    issn = {16155262}
}

@article{Krishnan2019SCADAForensics,
    title = {{SCADA testbed for vulnerability assessments, penetration testing and incident forensics}},
    year = {2019},
    journal = {7th International Symposium on Digital Forensics and Security, ISDFS 2019},
    author = {Krishnan, Sundar and Wei, Mingkui},
    month = {6},
    publisher = {Institute of Electrical and Electronics Engineers Inc.},
    isbn = {9781728128276},
    doi = {10.1109/ISDFS.2019.8757543}
}

@article{Ozcelik2021CENTERManagement,
    title = {{CENTER Water: A Secure Testbed Infrastructure Proposal for Waste and Potable Water Management}},
    year = {2021},
    journal = {9th International Symposium on Digital Forensics and Security, ISDFS 2021},
    author = {Ozcelik, Ibrahim and Iskefiyeli, Murat and Balta, Musa and Akpinar, Kevser Ovaz and Toker, Firdevs Sevde},
    month = {6},
    publisher = {Institute of Electrical and Electronics Engineers Inc.},
    isbn = {9781665444811},
    doi = {10.1109/ISDFS52919.2021.9486364}
}

@article{Steiner2021VWaterLabs:Education,
    title = {{vWaterLabs: developing hands-on laboratories for water-focused industrial control systems cybersecurity education}},
    year = {2021},
    journal = {Journal of Computing Sciences in Colleges},
    author = {Steiner, Stu and Kirkland, Matthew J and Conte De Leon, Daniel},
    month = {4},
    publisher = {Consortium for Computing Sciences in CollegesPUB305},
    url = {https://dl-acm-org.kfupm.idm.oclc.org/doi/10.5555/3470215.3470218},
    doi = {10.5555/3470215.3470218}
}

@article{Giehl2024EmuFlex:UA,
    title = {{EmuFlex: A Flexible OT Testbed for Security Experiments with OPC UA}},
    year = {2024},
    journal = {ACM International Conference Proceeding Series},
    author = {Giehl, Alexander and Heinl, Michael P. and Embacher, Victor},
    month = {7},
    volume = {1},
    publisher = {Association for Computing Machinery},
    url = {https://dl-acm-org.kfupm.idm.oclc.org/doi/10.1145/3664476.3670931},
    isbn = {9798400717185},
    doi = {10.1145/3664476.3670931}
}

@article{Ahmed2016APedagogy,
    title = {{A SCADA system testbed for cybersecurity and forensic research and pedagogy}},
    year = {2016},
    journal = {ACM International Conference Proceeding Series},
    author = {Ahmed, Irfan and Roussev, Vassil and Johnson, William and Senthivel, Saranyan and Sudhakaran, Sneha},
    month = {12},
    pages = {1--9},
    publisher = {Association for Computing Machinery},
    url = {https://dl-acm-org.kfupm.idm.oclc.org/doi/10.1145/3018981.3018984},
    isbn = {9781450347884},
    doi = {10.1145/3018981.3018984}
}

@article{Boeding2023AImplementations,
    title = {{A Testbed for Evaluating Performance and Cybersecurity Implications of IEC-61850 GOOSE Hardware Implementations}},
    year = {2023},
    journal = {Proceedings - IEEE Consumer Communications and Networking Conference, CCNC},
    author = {Boeding, Matthew and Hempel, Michael and Sharif, Hamid and Lopez, Juan and Perumalla, Kalyan},
    volume = {2023-January},
    publisher = {Institute of Electrical and Electronics Engineers Inc.},
    isbn = {9781665497343},
    doi = {10.1109/CCNC51644.2023.10060534},
    issn = {23319860}
}

@article{Green2017PainsResearch,
    title = {{Pains, Gains and PLCs: Ten Lessons from Building an Industrial Control Systems Testbed for Security Research}},
    year = {2017},
    journal = {CSET @ USENIX Security Symposium},
    author = {Green, B. and Le, Anhtuan and Antrobus, Rob and Roedig, U. and Hutchison, D. and Rashid, A.}
}

@article{Smadi2021AChallenges,
    title = {{A Comprehensive Survey on Cyber-Physical Smart Grid Testbed Architectures: Requirements and Challenges}},
    year = {2021},
    journal = {Electronics},
    author = {Smadi, Abdallah A. and Ajao, Babatunde Tobi and Johnson, Brian K. and Lei, Hangtian and Chakhchoukh, Yacine and Al-Haija, Qasem Abu},
    number = {9},
    month = {5},
    volume = {10},
    publisher = {Multidisciplinary Digital Publishing Institute (MDPI)},
    doi = {10.3390/ELECTRONICS10091043},
    issn = {20799292}
}

@article{Sankar2025DigitalControl,
    title = {{Digital Twin-Based Real-Time Industrial Monitoring and Control}},
    year = {2025},
    journal = {International Conference on Computing, Intelligence, and Application, CIACON 2025},
    author = {Sankar, P. and Vallikannu, R. and Charulatha, Siva and Kumar, Pratyush and Kumar, P. Anil and Kumar, Suraj},
    publisher = {Institute of Electrical and Electronics Engineers Inc.},
    isbn = {9798331544119},
    doi = {10.1109/CIACON65473.2025.11189434}
}

@article{Varghese2022DigitalSystems,
    title = {{Digital Twin-based Intrusion Detection for Industrial Control Systems}},
    year = {2022},
    journal = {2022 IEEE International Conference on Pervasive Computing and Communications Workshops and other Affiliated Events, PerCom Workshops 2022},
    author = {Varghese, Seba Anna and Dehlaghi Ghadim, Alireza and Balador, Ali and Alimadadi, Zahra and Papadimitratos, Panos},
    pages = {611--617},
    publisher = {Institute of Electrical and Electronics Engineers Inc.},
    isbn = {9781665416474},
    doi = {10.1109/PERCOMWORKSHOPS53856.2022.9767492},
    arxivId = {2207.09999}
}

@article{Bamunuarachchi2021DigitalTransformation,
    title = {{Digital Twins Supporting Efficient Digital Industrial Transformation}},
    year = {2021},
    journal = {Sensors 2021, Vol. 21, Page 6829},
    author = {Bamunuarachchi, Dinithi and Georgakopoulos, Dimitrios and Banerjee, Abhik and Jayaraman, Prem Prakash},
    number = {20},
    month = {10},
    pages = {6829},
    volume = {21},
    publisher = {Multidisciplinary Digital Publishing Institute},
    url = {https://www.mdpi.com/1424-8220/21/20/6829/htm https://www.mdpi.com/1424-8220/21/20/6829},
    doi = {10.3390/S21206829},
    issn = {1424-8220},
    pmid = {34696042}
}

@article{Nguyen2022DigitalTool,
    title = {{Digital Twin for IoT Environments: A Testing and Simulation Tool}},
    year = {2022},
    journal = {Communications in Computer and Information Science},
    author = {Nguyen, Luong and Segovia, Mariana and Mallouli, Wissam and Oca, Edgardo Montes de and Cavalli, Ana R.},
    pages = {205--219},
    volume = {1621 CCIS},
    publisher = {Springer Science and Business Media Deutschland GmbH},
    url = {https://link.springer.com/chapter/10.1007/978-3-031-14179-9_14},
    isbn = {9783031141782},
    doi = {10.1007/978-3-031-14179-9{\_}14/SAVE-RESEARCH},
    issn = {18650937}
}

@article{Ortiz2025Digital4.0,
    title = {{Digital Twin-Based Active Learning for Industrial Process Control and Supervision in Industry 4.0}},
    year = {2025},
    journal = {Sensors 2025, Vol. 25, Page 2076},
    author = {Ortiz, Jessica S. and Quishpe, Evelin K. and Sailema, Grace X. and Guam{\'{a}}n, Nathaly S.},
    number = {7},
    month = {3},
    pages = {2076},
    volume = {25},
    publisher = {Multidisciplinary Digital Publishing Institute},
    url = {https://www.mdpi.com/1424-8220/25/7/2076/htm https://www.mdpi.com/1424-8220/25/7/2076},
    doi = {10.3390/S25072076},
    issn = {1424-8220},
    pmid = {40218590}
}

@article{Dietz2020UnleashingSecurity,
    title = {{Unleashing the Digital Twin's Potential for ICS Security}},
    year = {2020},
    journal = {IEEE Security and Privacy},
    author = {Dietz, Marietheres and Pernul, Gunther},
    number = {4},
    month = {7},
    pages = {20--27},
    volume = {18},
    publisher = {Institute of Electrical and Electronics Engineers Inc.},
    doi = {10.1109/MSEC.2019.2961650},
    issn = {15584046}
}

@article{Alcaraz2022DigitalThreats,
    title = {{Digital Twin: A Comprehensive Survey of Security Threats}},
    year = {2022},
    journal = {IEEE Communications Surveys and Tutorials},
    author = {Alcaraz, Cristina and Lopez, Javier},
    number = {3},
    month = {9},
    pages = {1475--1503},
    volume = {24},
    publisher = {Institute of Electrical and Electronics Engineers Inc.},
    doi = {10.1109/COMST.2022.3171465},
    issn = {1553877X}
}

@article{Abraham2025TowardsApproaches,
    title = {{Towards enhanced cybersecurity in industrial control systems: a systematic review of context-based modeling, digital twins, and machine learning approaches}},
    year = {2025},
    journal = {International Journal of Information Security},
    author = {Abraham, Doney and Erceylan, Gizem and Gkioulos, Vasileios and Houmb, Siv Hilde},
    number = {6},
    month = {12},
    pages = {243-},
    volume = {24},
    publisher = {Springer Science and Business Media Deutschland GmbH},
    url = {https://link.springer.com/article/10.1007/s10207-025-01158-1},
    doi = {10.1007/S10207-025-01158-1/FIGURES/3},
    issn = {16155270}
}

@article{Empl2025DigitalPerspectives,
    title = {{Digital Twins in Security Operations: State of the Art and Future Perspectives}},
    year = {2025},
    journal = {ACM Computing Surveys},
    author = {Empl, Philip and Koch, David and Dietz, Marietheres and Pernul, Günther},
    number = {1},
    month = {2},
    volume = {58},
    publisher = {Association for Computing Machinery},
    url = {https://dl.acm.org/doi/10.1145/3746279},
    doi = {10.1145/3746279/ASSET/5824448E-6054-46B6-A9B0-70AA4CEBAC16/ASSETS/IMAGES/LARGE/CSUR-2024-0158-T6B.JPG},
    issn = {15577341}
}

@article{Ekisa2024VirtualApplications,
    title = {{Virtual Industrial Control System (ICS) Testbeds: A Comprehensive Review of Design, Implementation, and Cybersecurity Applications}},
    year = {2024},
    journal = {2024 Cyber Research Conference - Ireland, Cyber-RCI 2024},
    author = {Ekisa, Conrad and Briain, Diarmuid and Kavanagh, Yvonne},
    publisher = {Institute of Electrical and Electronics Engineers Inc.},
    isbn = {9798350390100},
    doi = {10.1109/CYBER-RCI60769.2024.10939771}
}

@article{Allison2023DigitalSystems,
    title = {{Digital Twin-Enhanced Incident Response for Cyber-Physical Systems}},
    year = {2023},
    journal = {ACM International Conference Proceeding Series},
    author = {Allison, David and Smith, Paul and McLaughlin, Kieran},
    month = {8},
    volume = {1},
    publisher = {Association for Computing Machinery},
    url = {https://dl.acm.org/doi/10.1145/3600160.3600195},
    isbn = {9798400707728},
    doi = {10.1145/3600160.3600195}
}

@article{AkereleAbimbola2023TheInfrastructure,
    title = {{The Digital Twins Incident Response To Improve the Security of Power System Critical Infrastructure}},
    year = {2023},
    journal = {Journal of Computing Sciences in Colleges},
    author = {{AkereleAbimbola} and {LeppertWilliam} and {SomervilleShionta} and {AmoussouGuy-Alain}},
    month = {10},
    publisher = {Consortium for Computing Sciences in CollegesPUB305},
    url = {https://dl.acm.org/doi/10.5555/3636988.3637008},
    doi = {10.5555/3636988.3637008}
}

@article{Allison2022DigitalApplications,
    title = {{Digital Twin-Enhanced Methodology for Training Edge-Based Models for Cyber Security Applications}},
    year = {2022},
    journal = {IEEE International Conference on Industrial Informatics (INDIN)},
    author = {Allison, David and Smith, Paul and McLaughlin, Kieran},
    pages = {226--232},
    volume = {2022-July},
    publisher = {Institute of Electrical and Electronics Engineers Inc.},
    isbn = {9781728175683},
    doi = {10.1109/INDIN51773.2022.9976095},
    issn = {19354576}
}

@article{Atalay2020AStandardization,
    title = {{A Digital Twins Approach to Smart Grid Security Testing and Standardization}},
    year = {2020},
    journal = {2020 IEEE International Workshop on Metrology for Industry 4.0 and IoT, MetroInd 4.0 and IoT 2020 - Proceedings},
    author = {Atalay, Manolya and Angin, Pelin},
    month = {6},
    pages = {435--440},
    publisher = {Institute of Electrical and Electronics Engineers Inc.},
    isbn = {9781728148922},
    doi = {10.1109/METROIND4.0IOT48571.2020.9138264}
}

@article{Akbarian2023DetectingTwins,
    title = {{Detecting and Mitigating Actuator Attacks on Cloud Control Systems through Digital Twins}},
    year = {2023},
    journal = {2023 31st International Conference on Software, Telecommunications and Computer Networks, SoftCOM 2023},
    author = {Akbarian, Fatemeh and Tarneberg, William and Fitzgerald, Emma and Kihl, Maria},
    publisher = {Institute of Electrical and Electronics Engineers Inc.},
    isbn = {9798350301076},
    doi = {10.23919/SOFTCOM58365.2023.10271648}
}

@article{Ayyalusamy2022HybridAnalysis,
    title = {{Hybrid Digital Twin Architecture for Power System Cyber Security Analysis}},
    year = {2022},
    journal = {Proceedings of the 11th International Conference on Innovative Smart Grid Technologies - Asia, ISGT-Asia 2022},
    author = {Ayyalusamy, V. and Sivaneasan, B. and Kandasamy, N. K. and Xiao, J. F. and Abidi, K. and Chandra, A.},
    pages = {270--274},
    publisher = {Institute of Electrical and Electronics Engineers Inc.},
    isbn = {9798350399660},
    doi = {10.1109/ISGTASIA54193.2022.10003563}
}

@article{Dauphinais2023AutomatedSystems,
    title = {{Automated Vulnerability Testing and Detection Digital Twin Framework for 5G Systems}},
    year = {2023},
    journal = {2023 IEEE 9th International Conference on Network Softwarization: Boosting Future Networks through Advanced Softwarization, NetSoft 2023 - Proceedings},
    author = {Dauphinais, Danielle and Zylka, Michael and Spahic, Harris and Shaik, Farhan and Yang, Jingda and Cruz, Isabella and Gibson, Jakob and Wang, Ying},
    pages = {308--310},
    publisher = {Institute of Electrical and Electronics Engineers Inc.},
    isbn = {9798350399806},
    doi = {10.1109/NETSOFT57336.2023.10175496}
}

@article{Dietz2020IntegratingCenter,
    title = {{Integrating digital twin security simulations in the security operations center}},
    year = {2020},
    journal = {ACM International Conference Proceeding Series},
    author = {Dietz, Marietheres and Vielberth, Manfred and Pernul, Günther},
    month = {8},
    publisher = {Association for Computing Machinery},
    url = {https://dl.acm.org/doi/10.1145/3407023.3407039},
    isbn = {9781450388337},
    doi = {10.1145/3407023.3407039}
}

@article{Grasselli2022AnSystems,
    title = {{An Industrial Network Digital Twin for enhanced security of Cyber-Physical Systems}},
    year = {2022},
    journal = {2022 International Symposium on Networks, Computers and Communications, ISNCC 2022},
    author = {Grasselli, Chiara and Melis, Andrea and Rinieri, Lorenzo and Berardi, Davide and Gori, Giacomo and Sadi, Amir Al},
    publisher = {Institute of Electrical and Electronics Engineers Inc.},
    isbn = {9781665485449},
    doi = {10.1109/ISNCC55209.2022.9851731}
}

@article{Gaurav2023ALearning,
    title = {{A DDoS Attack Detection System for Industry 5.0 using Digital Twins and Machine Learning}},
    year = {2023},
    journal = {GCCE 2023 - 2023 IEEE 12th Global Conference on Consumer Electronics},
    author = {Gaurav, Akshat and Gupta, Brij B. and Tai Chui, Kwok and Arya, Varsha and Benkhelifa, Elhadj},
    pages = {1019--1022},
    publisher = {Institute of Electrical and Electronics Engineers Inc.},
    isbn = {9798350340181},
    doi = {10.1109/GCCE59613.2023.10315663}
}

@inproceedings{Hammoudeh2020BlockchainInfrastructure,
    title     = {{Blockchain, Internet of Things and Digital Twins in Trustless Security of Critical National Infrastructure}},
    author    = {Hammoudeh, Mohammad},
    year      = {2020},
    booktitle = {Proceedings of the 4th International Conference on Future Networks and Distributed Systems},
    pages     = {1--1},
    publisher = {Association for Computing Machinery (ACM)},
    isbn      = {9781450388863},
    doi       = {10.1145/3440749.3442650},
    url       = {https://dl.acm.org/doi/10.1145/3440749.3442650}
}

@article{Lin2023DT4I4-Secure:Security,
    title = {{DT4I4-Secure: Digital Twin Framework for Industry 4.0 Systems Security}},
    year = {2023},
    journal = {2023 IEEE 14th Annual Ubiquitous Computing, Electronics and Mobile Communication Conference, UEMCON 2023},
    author = {Lin, Yu Zheng and Shao, Sicong and Rahman, Md Habibor and Shafae, Mohammed and Satam, Pratik},
    pages = {200--209},
    publisher = {Institute of Electrical and Electronics Engineers Inc.},
    isbn = {9798350304138},
    doi = {10.1109/UEMCON59035.2023.10316090}
}

@article{Balta2024DigitalSystems,
    title = {{Digital Twin-Based Cyber-Attack Detection Framework for Cyber-Physical Manufacturing Systems}},
    year = {2024},
    journal = {IEEE Transactions on Automation Science and Engineering},
    author = {Balta, Efe C. and Pease, Michael and Moyne, James and Barton, Kira and Tilbury, Dawn M.},
    number = {2},
    month = {4},
    pages = {1695--1712},
    volume = {21},
    publisher = {Institute of Electrical and Electronics Engineers Inc.},
    doi = {10.1109/TASE.2023.3243147},
    issn = {15583783}
}

@article{Chukkapalli2021Cyber-PhysicalUsecase,
    title = {{Cyber-Physical System Security Surveillance using Knowledge Graph based Digital Twins - A Smart Farming Usecase}},
    year = {2021},
    journal = {Proceedings - 2021 IEEE International Conference on Intelligence and Security Informatics, ISI 2021},
    author = {Chukkapalli, Sai Sree Laya and Pillai, Nisha and Mittal, Sudip and Joshi, Anupam},
    publisher = {Institute of Electrical and Electronics Engineers Inc.},
    isbn = {9781665438384},
    doi = {10.1109/ISI53945.2021.9624688}
}
\end{document}